\pdfoutput=1
\documentclass[cernpreprint,USenglish]{na61doc}

\usepackage{colortbl}
\definecolor{darkred}{rgb}{0.5,0,0}
\definecolor{darkblue}{rgb}{0,0,0.5}
\definecolor{firebrick}{rgb}{0.75,0.125,0.125}
\definecolor{darkgreen}{rgb}{0,0.5,0}

\usepackage[colorlinks=true,linkcolor=firebrick,citecolor=darkgreen,urlcolor=darkblue]{hyperref}

\usepackage{myphi}

\ShineTitle{Measurement of \phi meson production in \pp interactions at 40, 80
  and \SI{158}{\GeVc} with the \NASixtyOne spectrometer at the CERN SPS}

\PreprintIdNumber{CERN-EP-2019-165}

\ShineJournal{Eur.\ Phys.\ J.\ C}
\ShineAbstract{ %
  Results on \phi meson production in inelastic \pp collisions at CERN SPS
energies are presented. They are derived from data collected by the \NASixtyOne
fixed target experiment, by means of invariant mass spectra fits in the $\phi
\to \Kp\Km$ decay channel. They include the first ever measured double differential
spectra of \phi mesons as a function of rapidity \y and transverse momentum \pt
for proton beam momenta of \SI{80}{\GeVc} and \SI{158}{\GeVc}, as well as
single differential spectra of \y or \pt for beam momentum of \SI{40}{\GeVc}.
The corresponding total \phi yields per inelastic \pp event are obtained. These
results are compared with existing data on \phi meson production in \pp
collisions. The comparison shows consistency but superior accuracy of the
present measurements. The emission of \phi mesons in \pp reactions is confronted
with that occurring in \PbPb collisions, and the experimental results are
compared with model predictions. It appears that none of the considered models
can properly describe all the experimental observables.

}

\begin{document}
\maketitle


\NewMathSymbol{\sigmaKp}{\sigma_\Kp}
\NewMathSymbol{\sigmaKm}{\sigma_\Km}
\NewMathSymbol{\sigmaPhi}{\sigma_\phi}
\NewMathSymbol{\totxsect}{\sigma_\text{tot}}
\NewMathSymbol{\elxsect}{\sigma_\text{el}}
\section{Introduction}
The motivation for studying particle production in proton-proton collisions is twofold. Firstly, such data are necessary to characterize soft hadronic interactions and to develop phenomenological models which are then used to describe the observable final states. Particle yields (only) are generally well described by statistical particle production models, see e.g.\ \recite{bib:Vovchenko2016}, while complete particle spectra are computed in e.g.\ microscopic (string) models~\cite{bib:pythia, bib:DPM, bib:Fritiof}. Secondly, they are considered as a trivial reference in the search for collective effects in heavy ion collisions at moderate energies. In this context the \phi meson is one of the most interesting hadrons, because  it consists of an $s$ and an $\bar{s}$ valence quark with only small admixtures of light valence quarks. Its net strangeness vanishes, which means that in a pure hadron scenario, \phi production is insensitive to strangeness-related effects. On the other hand, if partonic degrees of freedom are significant, the \phi will behave like a doubly-strange particle.
Therefore \phi mesons are expected to play a key role in studies of phenomena related to the phase transition separating the confined hadron and deconfined parton phase, the quark-gluon plasma. The transition is considered to occur in heavy ion collisions in the lower CERN SPS energy regime~\cite{bib:Onset2011}. 
Such parton matter may (can) be detected in the final state of nuclear collisions by studying the onset of medium effects which cannot be explained by hadron processes. Doubly-strange hadrons are considered to be sensitive to those medium effects. Thus the results on \phi production at beam momenta of \SI{40}{\GeVc}, \SI{80}{\GeVc}, and \SI{158}{\GeVc} presented in this paper serve as a pure hadron scenario reference for the comparison with results measured in nuclear collisions at the same energy. 
\par
Production of \phi mesons has been measured in colliding systems ranging from \ee to \PbPb reactions, and at energies from GSI SIS to CERN LHC accelerators.
In this paper double differential yields of \phi mesons produced in proton-proton collisions at \SI{80}{\GeVc} and \SI{158}{\GeVc} as well as single differential yields at \SI{40}{\GeVc} are presented and compared with published experimental data on \pp interactions \cite{bib:Blobel_pp24_phi_1975,
bib:ACCMOR_hadrons63_93_phi_1981, bib:Drijard_ppS53_phi_1981,
bib:AguilarBenitez_pp400_phi_1991, bib:NA49phi2000, bib:ANKE_pp2_3_phi_2008,
bib:STAR_many_reactions_phi_2009, bib:PHENIX_many_systems_sNN200_phi_2009,
bib:ALICE_phi_0.9TeV, bib:ALICE_phi_7TeV, bib:ATLAS_phi_7TeV,
bib:LHCb_phi_7TeV}, and on \PbPb collisions at the same energy~\cite{bib:NA49phi2008}.  For \pp collisions, measurements
exist of differential and total inclusive cross-sections at CERN SPS and ISR energies~\cite{bib:Blobel_pp24_phi_1975, bib:ACCMOR_hadrons63_93_phi_1981,
bib:Drijard_ppS53_phi_1981, bib:AguilarBenitez_pp400_phi_1991}. The NA49
collaboration published single differential spectra of rapidity and transverse momentum at the incoming beam energy of \SI{158}{\GeV}~\cite{bib:NA49phi2000}, allowing for
direct comparison with the present work. At higher collision energies
mainly the midrapidity region of \phi production is known
experimentally~\cite{bib:STAR_many_reactions_phi_2009,
bib:PHENIX_many_systems_sNN200_phi_2009, bib:ALICE_phi_0.9TeV,
bib:ALICE_phi_7TeV, bib:ATLAS_phi_7TeV}, with the exception of double
differential cross-sections measured in the forward region by the LHCb
experiment~\cite{bib:LHCb_phi_7TeV}.
\par
For the purpose of the comparison between \pp and \PbPb reactions, the present analysis
operates on multiplicities of \phi mesons produced per inelastic \pp collision rather than cross-sections. Note that the latter can be transformed into the former using tables of total (\totxsect) and elastic (\elxsect) proton-proton cross-sections as a function of collision energy~\cite{bib:PDGxsect}:
\begin{equation}
  n = \frac{\sigma}{\totxsect - \elxsect} \,,
  \label{eq:xsectConversion}
\end{equation}
where $n$ is the multiplicity per inelastic interaction while
$\sigma$ is the cross-section for \phi production.
\par
This paper is the fourth in a series of the \NASixtyOne collaboration presenting experimental results on particle production in \pp interactions at CERN SPS energies. The relevant details of beam, target, experimental setup, and event selection were already described in previous publications~\cite{bib:NA61_facility, bib:NA61_pC31_pion_2011, bib:NA61_pp_pion_2014}.
Therefore \secref{s:experiment} contains only a short description of the \NASixtyOne spectrometer, of the data samples, and of the event selection.  
\secref{s:analysis} summarizes the data analysis and systematic errors. \secref{s:Results_paper} presents and discusses the results of the present analysis together with the world data on \phi production in \pp and \PbPb collisions and compares them with calculations of the three microscopic models \Pythia, \EPOS, and \UrQMD~\cite{bib:pythia, bib:EPOS2006, bib:EPOS2009, bib:UrQMD1998, bib:UrQMD1999}. The latter two are also designed to describe nuclear collisions. A summary in \secref{s:summary} closes the paper.

The following variables and definitions are used in this paper. The particle rapidity \y is calculated
in the collision center of mass system (cms), $\y = 0.5 \ln[(E+c\pL)/(E-c\pL)]$, where $E$
and \pL are the particle energy and longitudinal momentum, respectively. The transverse component
of the momentum is denoted as \pt and the transverse mass \mt is defined as
$\mt = \sqrt{m^2 + (\pt / c)^2}$
where $m$ is the particle mass. The total momentum in the laboratory frame is denoted $p$ and the
collision energy per nucleon pair in the center of mass by \sNN.

\begin{figure*}[t!]
  \centering
  \includegraphics[width=\textwidth]{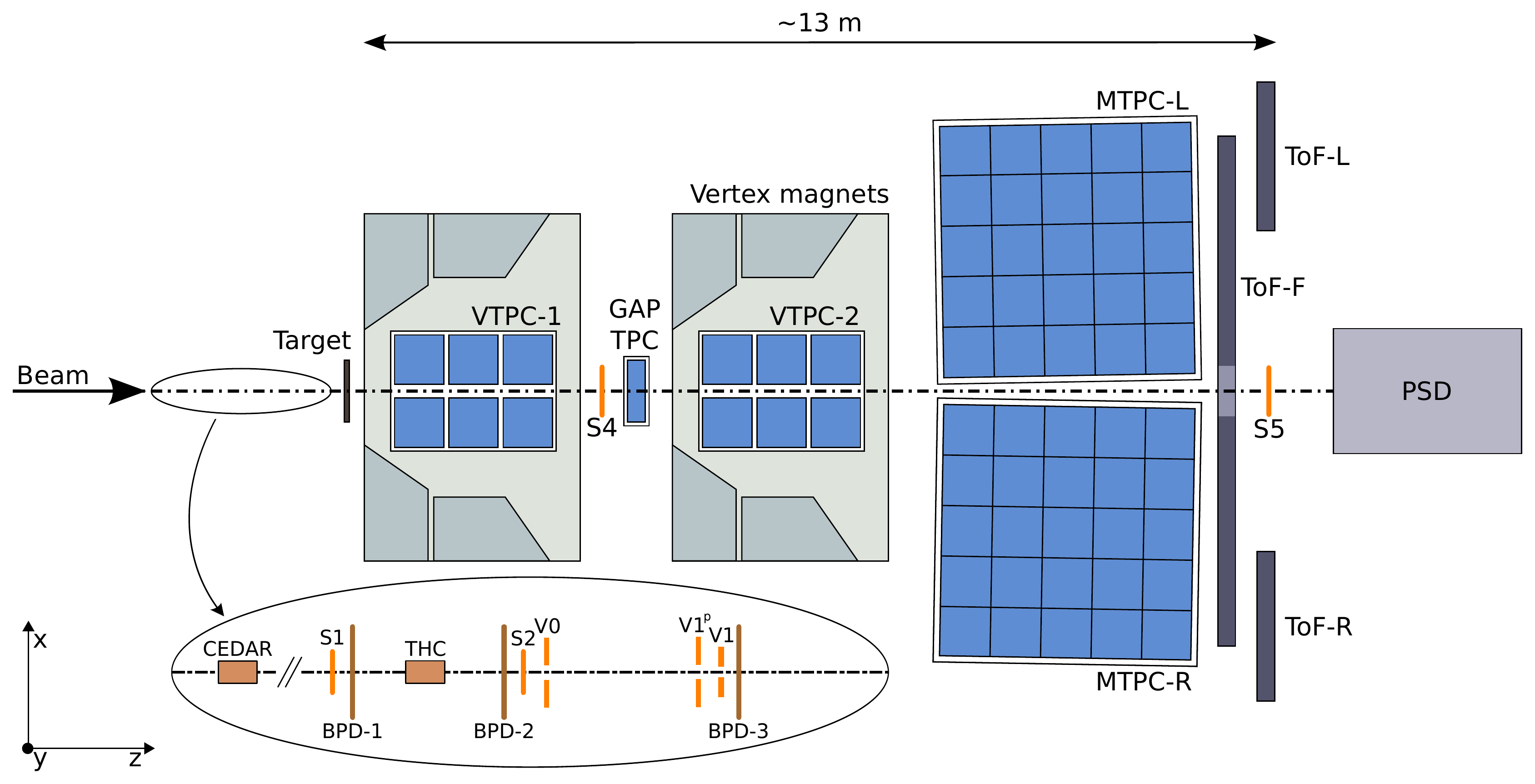}
  \caption{
    Schematic layout of the \NASixtyOne detector system (horizontal cut in the
    beam plane, not to scale). Also outlined are the coordinate system used in
    the experiment and the beam detector configuration used with secondary proton
    beams in 2009.
  }
  \label{fig:setup}
\end{figure*}
\section{The \NASixtyOne experiment}\label{s:experiment}
\NASixtyOne is a fixed target experiment conducted in the North Area of the
CERN Super Proton Synchrotron (SPS) accelerator complex. The detector system of
\NASixtyOne, depicted in \figref{fig:setup}, is described in detail in
\recite{bib:NA61_facility}. 
The data studied in the present analysis were collected with secondary beams of positively charged hadrons at 40, 80 and \SI{158}{\GeVc}. The latter were produced by \SI{400}{\GeVc}
protons extracted from the SPS onto a beryllium target in the slow extraction mode 			
with a flat-top of 10 seconds.
The secondary beam momentum and intensity was adjusted by proper setting of
the H2 beam-line magnet currents and collimators. The beam was transported along the H2 beam-line towards the experiment.
The precision of the bending power of the beam magnets was approximately 0.5\%.
The protons in the secondary hadron beam (58\% at \SI{158}{\GeVc}, 28\% at \SI{75}{\GeVc} and 14\% at \SI{40}{\GeVc}) 
were identified by two Cherenkov counters, a CEDAR (either \mbox{CEDAR-W} or \mbox{CEDAR-N}) and a threshold counter (THC). The CEDAR counter, using a coincidence of six out
of the eight photo-multipliers placed radially along the Cherenkov ring,
provided positive identification of protons, while the THC, operated at pressure 
lower than the proton threshold, was used in anti-coincidence in
the trigger logic. Due to their limited range of operation two different
CEDAR counters were employed, namely for beams at 20, 31, and \SI{40}{\GeVc} the \mbox{CEDAR-W} counter and
for beams at 80 and \SI{158}{\GeVc} the \mbox{CEDAR-N} counter. The threshold counter
was used for all beam energies. This scheme allowed to select beam protons with a purity of about 
99\%. Beam particle trajectories were measured by a set of three beam position detectors (BPDs) used to determine the transverse position of the collision point. The beam trigger used the information from plastic scintillator and Cherenkov counters. The interaction trigger consisted of
the beam trigger and a veto-signal from a \SI{2}{\cm} diameter scintillator (S4) placed approximately \SI{4}{\m} downstream from the target on the trajectory of the beam. This minimum bias trigger required that a valid beam proton is absent downstream of the target. There was, however, a non-negligible probability that a charged particle from an inelastic collision hits S4 and inhibits the recording of the associated event. This bias is taken into account by a Monte Carlo correction.
The target was a liquid hydrogen vessel. It was a \SI{20.29}{\cm} long (\SI{2.8}{\percent} of nuclear interaction length) cylinder with a diameter of \SI{3}{\cm}. The liquid hydrogen had a density of approximately \SI{0.07}{g/cm^3}.
\par
The main components of the detection system used in the analysis are four large volume Time Projection Chambers (TPC). Two of them, called Vertex TPCs (VTPC), are located approximately \SI{80} cm downstream of the target centered inside superconducting magnets which provide a maximum combined bending power of 9 Tm. Two further TPCs (MTPC) are placed side by side in the field free region behind the magnets. The TPCs are filled with Ar:CO$_2$ gas mixtures in proportions 90:10 for the VTPCs and 95:5 for the Main TPCs.  Two walls of pixel Time-of-Flight (ToF-L/R) detectors are placed symmetrically to the beamline downstream of the Main TPCs. Each wall contains 891 individual scintillation detectors with rectangular dimensions, each having a single photomultiplier tube glued to the short side. The scintillators have a thickness of \SI{23}{\mm} matched to the photocathode diameter, a height of \SI{34}{\mm} and horizontal width of 60, 70 or \SI{80}{\mm}, with the shortest scintillators positioned closest to the beamline and the longest on the far end. A GAP-TPC (GTPC) between VTPC-1 and VTPC-2 improves the acceptance for high-momentum forward-going particles. The TPCs record the tracks and energy loss (\dEdx) of the charged particles produced in the collision. Their momentum vectors are calculated from the track parameters and the magnetic field.
\par
The present analysis was performed on minimum bias proton-proton collision data at three beam momenta 158, 80 and \SI{40}{\GeVc}.
The recorded and selected event statistics are shown in \tabref{tab:data}. The difference between the two numbers is caused by the event selection cuts (see below).

\begin{table}[bt]
  \centering
  \caption{Number of events recorded in 2009 
    and selected for the \phi analysis. 
  }
  \begin{tabular}{@{}c@{\qquad}c@{\qquad}c@{}}
    \toprule
    \pbeam [\si{\GeVc}] & recorded & selected  \\
    \midrule
    158 & 3.5$\cdot 10^6$ & 1.3$\cdot 10^6$ \\
    80 & 4.5$\cdot 10^6$ & 1.3$\cdot 10^6$ \\
    40 & 5.2$\cdot 10^6$ & 1.6$\cdot 10^6$ \\
     \bottomrule
  \end{tabular}
  \label{tab:data}
\end{table}
\par
A large sample of Monte Carlo (MC) events was generated in order to estimate the corrections for detector and analysis deficiencies. The MC samples contained 20 million \pp events at each
collision energy. These were generated using the \Epos~1.99
model~\cite{bib:EPOS2006,bib:EPOS2009} available within the \CRMC~1.4
package~\cite{bib:CRMC}. The detector response was simulated using the
\Geant~3.21 package~\cite{bib:Geant}. Event reconstruction was performed by the same \NASixtyOne software version as used for the treatment of experimental data. Two modifications were applied to the original \Epos code: the natural width of the \phi
resonance was adjusted to its PDG value~\cite{bib:PDG}; the branching
ratio for the $\phi\to\Kp\Km$ decay channel was set to \SI{100}{\percent} to
increase the number of detectable \phi decays. By virtue of the relatively small \phi multiplicity, this
latter change has no significant effect on the overall event characteristics and thus does not bias the obtained corrections. \Epos was chosen as event generator, because other tested
models performed worse in comparison with \NASixtyOne results on hadron
production in hadron-hadron and hadron-nucleus interactions
\cite{bib:IlnickaMSc,bib:UngerICHEP2010,bib:UngerISVHECRI2012,
bib:NA61_pC31_pion_2011}.
\par
Well studied cuts were applied to obtain a clean sample of inelastic \pp events (see \recite{bib:NA61_pp_pion_2014}). These include the requirements of the reconstruction of the interacting beam particle in the Beam Position Detectors and of the interaction point well inside the target vessel. Furthermore, events with a single, well measured positively charged track with absolute momentum close to the beam momentum were rejected. These are considered to be elastic events in which
the beam proton scattered elastically into the acceptance of the TPCs.
This rejection was needed only for the two lower beam momenta, because at \SI{158}{\GeVc} the veto counter intercepted essentially all of the forward going protons from elastic \pp interactions~\cite{bib:NA61_pp_pion_2014}.

\section{Analysis methodology}\label{s:analysis}
\NewMathSymbol{\PIDeffconstraint}{\PIDeff_\y}
\newcommand{\correction}[1]{c_\text{#1}}
\NewMathSymbol{\intCorr}{\correction{$\infty$}}
\NewMathSymbol{\mcCorr}{\correction{MC}}
\NewMathSymbol{\bkgCorr}{\correction{bkg}}
This section outlines the analysis procedure and describes the details of track selection, of \phi signal extraction as well as the necessary corrections and systematic uncertainties.
Since \phi mesons cannot be detected directly, they are identified using the most frequent charged particle decay mode $\phi\to\Kp\Km$. Their yield is obtained from the invariant mass distribution of pairs of oppositely charged particles assuming the kaon mass. Decays of \phi mesons into $\Kp$ and $\Km$ manifest themselves as a resonance signal on a background of uncorrelated pairs and correlated pairs from decays of other unstable particles or resonances into oppositely charged particles. The number of uncorrelated pairs is significantly reduced, if only charged kaons are considered. Therefore  
kaon candidates are selected using the information about particle momenta and energy loss provided by the TPCs, as well as time-of-flight provided by the TOF-walls. The resulting invariant mass spectrum contains correlated $\Kp$-$\Km$ pairs, correlated pairs of charged particles with one or two wrong mass assignments, and uncorrelated pairs. The significance of the \phi signal depends on the quality of the kaon identification, and the phase space distribution of the contributing particles. The number of \phi mesons is determined by fitting suitable parametrizations of the signal and of the background to the invariant mass distributions.   
\par
The trajectories of the charged particles (the tracks) used in the invariant mass analysis are reconstructed using TPC data. The reconstructed tracks are subjected to quality checks to select particles produced in the primary interaction, to ensure good momentum resolution, and to reduce fakes. For a complete description of the track cuts see \recite{bib:MarcinekPhD}.
Their distance of closest approach to the interaction point (main vertex) must not exceed 4 cm in the bend plane and 2 cm in the plane spanned by the beam and magnetic field direction. A further
criterion requires that the tracks consist of more than 30 clusters (\enquote{points}). This ensures reasonable \dEdx resolution. In addition the number of clusters per track reconstructed in the magnetic field must be larger than 15 in the VTPCs or more than 4 in the GAP TPC. This ensures reasonable momentum determination accuracy.
\begin{figure*}[t!]
  \centering
  \subfloat[]{
    \label{fig:PIDCut:band}
    \includegraphics[width=0.5\textwidth,page=2,clip,trim=0 5mm 0 0]{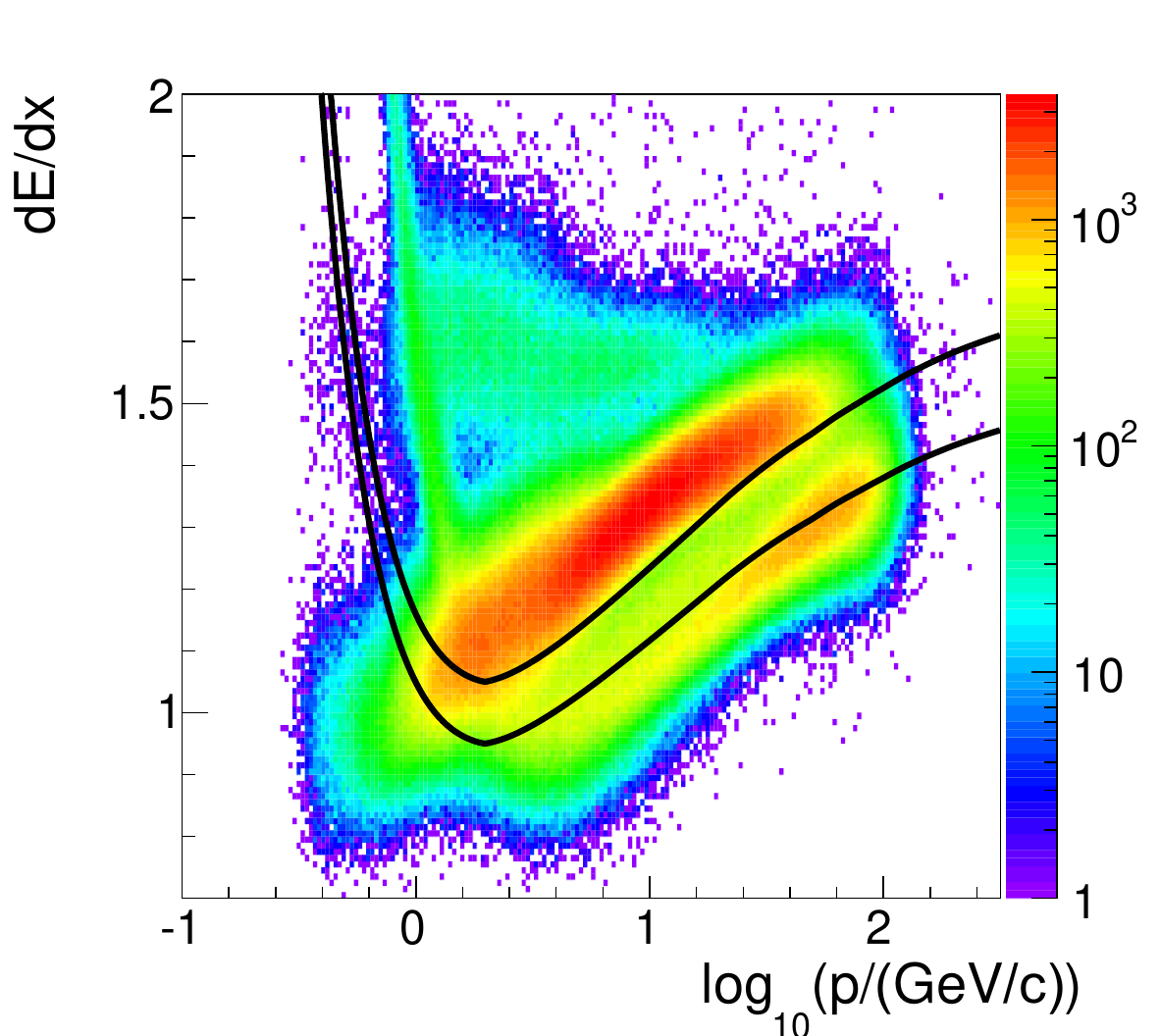}
  }
  \subfloat[]{
    \label{fig:PIDCut:TOF}
    \includegraphics[width=0.5\textwidth,page=2,clip,trim=0 0 134mm 0mm]{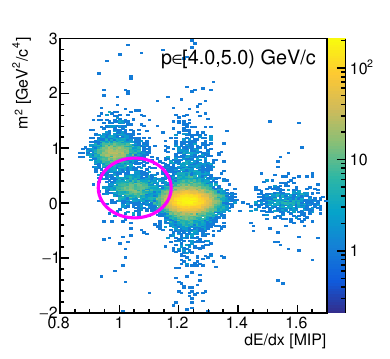}
  }
  \caption{Illustration of the kaon candidate selection. The data are
    from the \SI{158}{\GeVc} run. The band between the two black curves in
    panel~\subref{fig:PIDCut:band} is mainly populated by kaons and accepted in the analysis.  Structures associated with pions and protons are visible above and below
    the band.
    \subref{fig:PIDCut:TOF}~shows an example of how those particles are rejected which are with high probability not kaons by a cut in energy loss (\dEdx) and mass squared derived from TOF (outside of the pink circle).
  }
  \label{fig:PIDCut}
\end{figure*}
\par
The efficient selection of kaon candidates is of great importance for the \phi resonance analysis. It is mainly based on the momentum and energy loss  measurements along the trajectories of the charged particles in the TPCs. The correlation of both quantities for all accepted positively charged particles is shown in \figref{fig:PIDCut:band} in terms of their momenta and (truncated) mean energy losses $\dEdx$. Kaon candidates are selected by a momentum-dependent $\dEdx$ window around the expectation value. The size of this window was chosen such that the possible loss of kaons is small. This is achieved by selecting tracks with $\dEdx$ within $\pm\SI{5}{\percent}$ of the nominal $\dEdx$ curve as given by the Bethe-Bloch formula. The experimental $\dEdx$ resolution is roughly 5\%. The upper and lower limits of this cut are visualized as black lines 
in~\figref{fig:PIDCut:band}. Particle time-of-flight information is available near midrapidity and is used to reject those particles which are not a charged kaon. An example is shown in \figref{fig:PIDCut:TOF} where particles outside the pink circle are rejected. The details of the time-of-flight measurement and calibration were described in \recite{Aduszkiewicz:2017sei}.


\begin{figure*}[!t]
  \centering
  \subfloat[2D \ypt binning]{
    \label{fig:BinningTypes:2D}
    \includegraphics[width=0.5\textwidth,page=1]{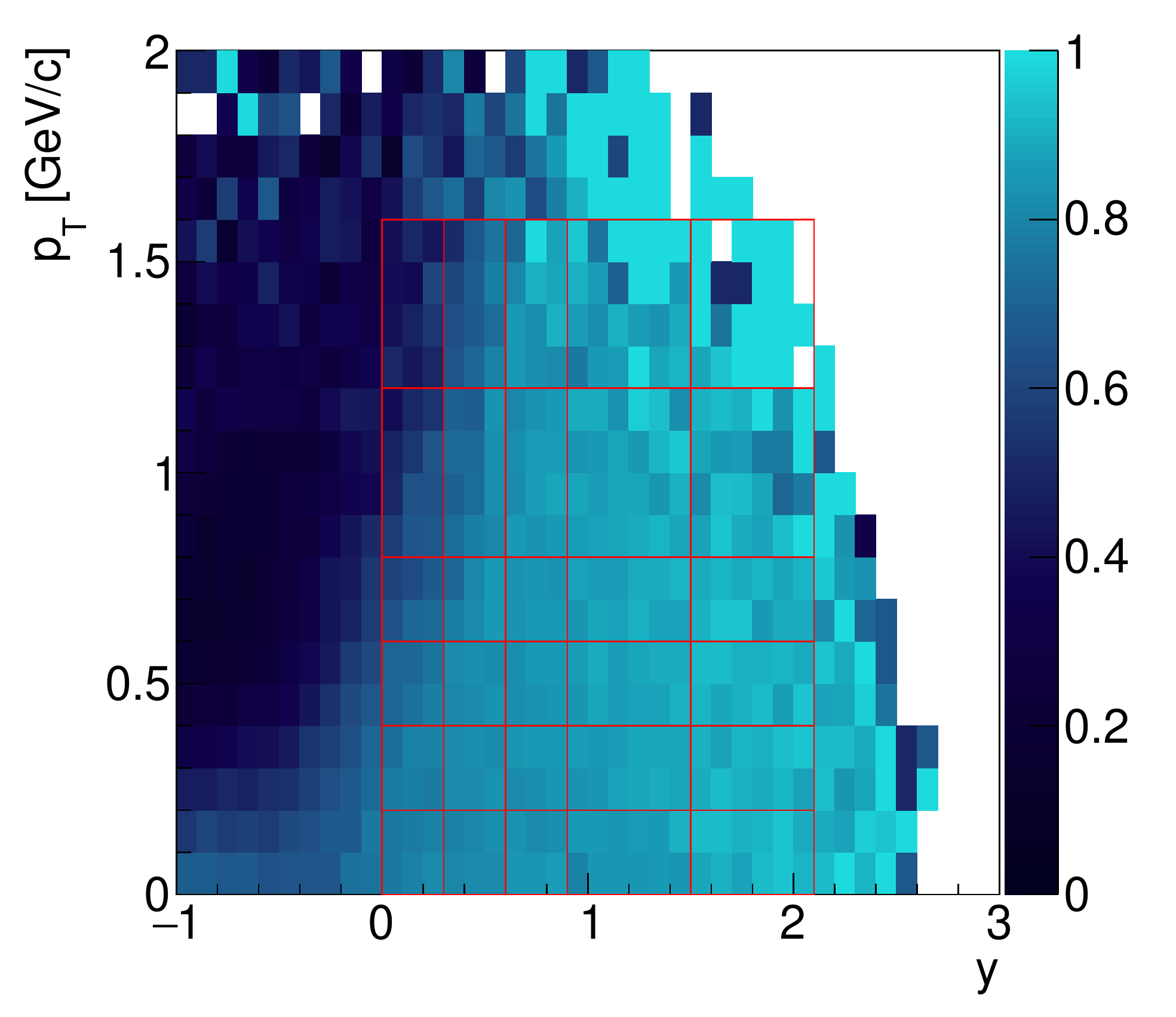}
  }
  \subfloat[1D \y binning]{
    \label{fig:BinningTypes:1Dy}
    \includegraphics[width=0.5\textwidth,page=2]{BinningTypes.pdf}
  }  \\ 
  \subfloat[1D \pt binning]{
    \label{fig:BinningTypes:1Dpt}
    \includegraphics[width=0.5\textwidth,page=6]{BinningTypes.pdf}
  }
  \subfloat[broad binning]{
    \label{fig:BinningTypes:broad}
    \includegraphics[width=0.5\textwidth,page=3]{BinningTypes.pdf}
  }
  \caption{Binning types used in this analysis, overlaid on the \phi
    registration probability obtained from simulations of inelastic p+p interactions at \SI{158}{\GeVc}. Empty regions correspond to bins where probability
    calculation was not possible due to insufficient statistics of generated
    particles.
  }
  \label{fig:BinningTypes}
\end{figure*}
The goal of the present analysis is to obtain the \phi meson production yields
in bins of rapidity \y and transverse momentum \pt. This requires the study of the invariant mass distributions for each considered \yptBin bin. Several
types of binning in rapidity and transverse momentum are used.
They are all illustrated in \figref{fig:BinningTypes}.  For comparison with other existing experimental data the results presented here are sometimes determined also in $(\y, \mtmRest)$ bins, where \restMass is the rest mass of the \phi meson. 
\par
\begin{figure*}[t]
  \centering
  \IfEPJC{
    \includegraphics[width=0.6\textwidth]{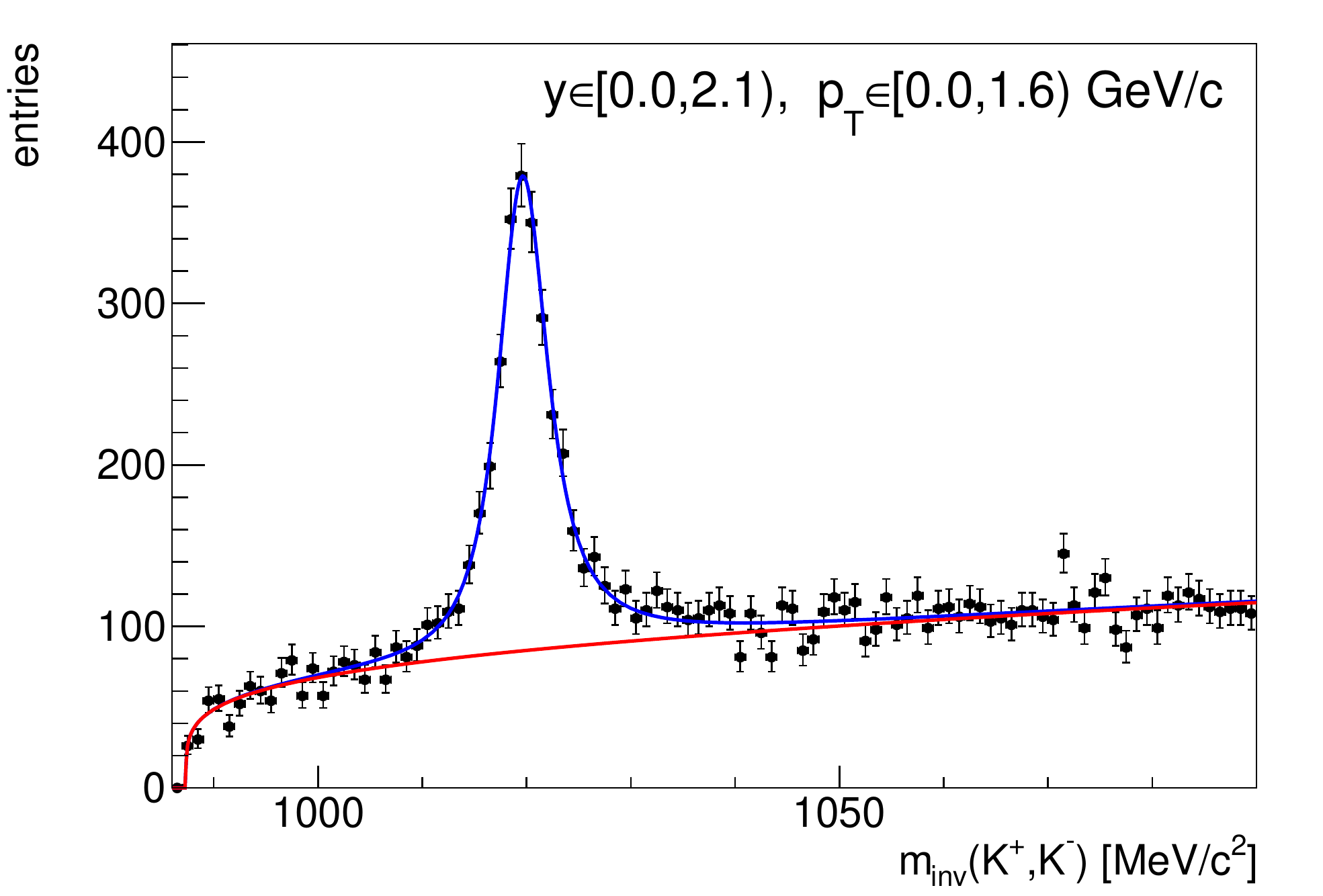}
  }{
    \includegraphics[width=0.7\textwidth]{14B026_pp158_unbinnedFit_Pt.pdf}
  }
  \caption{Example of a fitted invariant mass spectrum of kaon pair candidates in a large
    \phi phase space region as defined in \figref{fig:BinningTypes:broad}, obtained
    for inelastic p+p interactions at \SI{158}{\GeVc}. Both kaon candidates are subjected to the identification procedure. The signal shape parameters \mphi and $\sigma$
    resulting from this fit are used to constrain the fits in fine binned \phi phase space. The blue curve represents the fitted function defined by \equref{eq:SingleFitFunc}, while the red
    curve represents the background component. Its shape is given by the ARGUS function. See the text for details. 
  }
  \label{fig:SingleSpectrumFit}
\end{figure*}
\begin{figure*}[t]
  \centering
  \includegraphics[width=\textwidth,page=1]{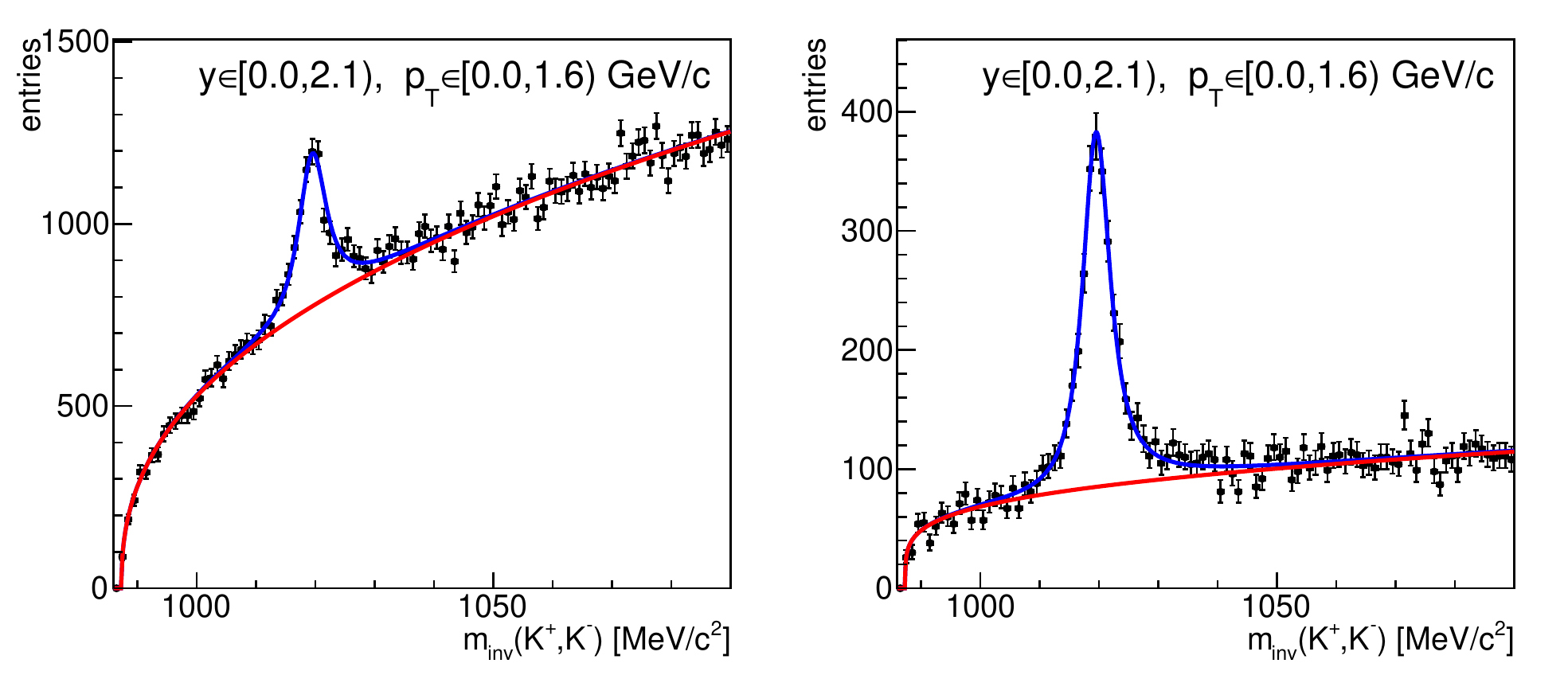}
  \caption{Illustration of a simultaneous tag-and-probe fit for the same data as shown in \figref{fig:SingleSpectrumFit} with only one of the kaon candidates subjected to the particle identification conditions (left) and same data as \figref{fig:SingleSpectrumFit} with both kaon candidates subjected to the PID procedures (right). Note that here the resonance signal parameters are kept fixed and the new parameter~\PIDeff is introduced. The blue curves represent the fitting function defined by \equref{eq:TagAndProbeFitFunc} and the red curves the background. See text for description of quoted parameters. 
  }
  \label{fig:TagAndProbeFit}
\end{figure*}
The invariant mass spectrum of \phi candidates in the \yptBin bin specified in \figref{fig:BinningTypes:broad} is shown in \figref{fig:SingleSpectrumFit}. The \phi~signal peaks around \SI{1020}{\MeVcsq} on a more or less structureless background.
The signal is parametrized with a function which contains two
components that take into account the natural shape of the resonance and its broadening due to the detector resolution. The first component
is described by a relativistic Breit-Wigner function:
\begin{subequations}
  \label{eq:RelBW}
  \begin{gather}
    L(x; \mphi, \Gamma) \propto
      \frac{x \Gamma_x(x)}{\pqty{x^2 - \mphi^2}^2 + \mphi^2 \Gamma_x^2(x)}
      \,, \label{eq:RelBW:main} \\
    \intertext{with}
    \Gamma_x(x) = 2 \Gamma \pqty{\frac{q(x)}{q(\mphi)}}^3
      \frac{q^2(\mphi)}{q^2(x) + q^2(\mphi)}
      \,, \label{eq:RelBW:gamma} \\
    \intertext{and}
    q(x) = \sqrt{\tfrac{1}{4}x^2 - m_K^2} \,,
  \end{gather}
\end{subequations}
where \mphi is the peak position (expected to be equal, within uncertainties, to the mass of the \phi meson), $\Gamma$ is the natural width of the \phi, and $m_K$ is the kaon mass. This parametrization was adopted from \recite{bib:NA49phi2008} and first introduced in \recite{bib:JDJackson}.
\par
The second component is described by the \qGaussian function:
\begin{equation}
  \label{eq:qGaus}
  G(x; \sigma, q) \propto
    \bqty{1 + (q - 1)\frac{x^2}{2\sigma^2}}^{-\frac{1}{q - 1}} \,,
\end{equation}
where $\sigma$ is the width and $q$ the shape parameter. The choice of
this parametrization is discussed in \recite{bib:MarcinekPhD}. As stated
there, the parameter $q$ is not fitted to data but fixed using a Monte Carlo study of the experimental invariant mass resolution. It no longer appears as a parameter of the function $G$.
\par
The resulting resonance peak function is given by the convolution of \equref{eq:RelBW} and \equref{eq:qGaus}:
\begin{equation}
  \label{eq:SignalParametrization}
  V(x; \mphi, \sigma, \Gamma) = L \ast G
    = \int_{-\infty}^{+\infty} G(x';\sigma) L(x-x'; \mphi, \Gamma) \dd{x'} \,.
\end{equation}
In practice, it is not possible to simultaneously fit the two width parameters, $\sigma$ and $\Gamma$. Therefore the $\Gamma$ parameter is fixed to its PDG value and dropped from the list of fitted parameters in all further equations.
\par

A reliable description of the background under the \phi signal must take into account that the signal is close to the lower kinematical limit of the invariant mass given by the mass of two kaons. 
We use the ARGUS function~\cite{bib:Argus1990} to describe the background under the \phi signal. The function has two shape parameters and reads:
\begin{subequations}
  \newcommand{\argusu}{\pqty{1 - \frac{z^2(x)}{x^2_\text{max}}}}
  \label{eq:argus}
  \begin{gather}
  B(x; k, p) =
  \begin{cases}
    0 & \text{for $x \leq 2m_K$} \\
    \IfEPJC{
      \begin{array}{@{}l@{}l@{}}
        z(x) & \cdot \argusu^p  \\
             & \cdot \exp{k \argusu}
      \end{array}
    }{
      z(x) \cdot \argusu^p \cdot \exp{k \argusu}
    }
      & \text{for $x > 2m_K$}
  \end{cases} \,, \\
  \intertext{with}
  z(x) = 2 m_K + x_\text{max} - x \,,
  \end{gather}
\end{subequations}
where $k$ is a shape parameter corresponding to $-\frac{1}{2}\chi^2$ in the
Wikipedia formula for the ARGUS distribution, $p$ is the power as in the
generalized ARGUS distribution, $m_K$ is the kaon mass and $x_\text{max}$ is
the right boundary of the \m histogram. Note that in this parametrization,
based on the class \texttt{RooArgusBG} from \cite{bib:RooFit},
$k$ can be any real number.
The complete function used to fit the invariant mass spectrum is shown as blue curve in \figref{fig:SingleSpectrumFit}. It is defined as:
\begin{equation}
  f(\m) =
    \Np V(\m; \mphi, \sigma) + \Nbkg B(\m;k,p) \,,
  \label{eq:SingleFitFunc}
\end{equation}
where $V(\m; \mphi, \sigma)$ is given by \equref{eq:SignalParametrization} and $B(\m;k,p)$ 
by \equref{eq:argus}. Both are normalised in such a way that \Np and \Nbkg are the number of signal and background pairs in the mass distribution. 

\subsection{The tag-and-probe method}\label{s:tagAndProbe}
\begin{figure*}[t]
  \centering
  \includegraphics[width=\textwidth,page=27]{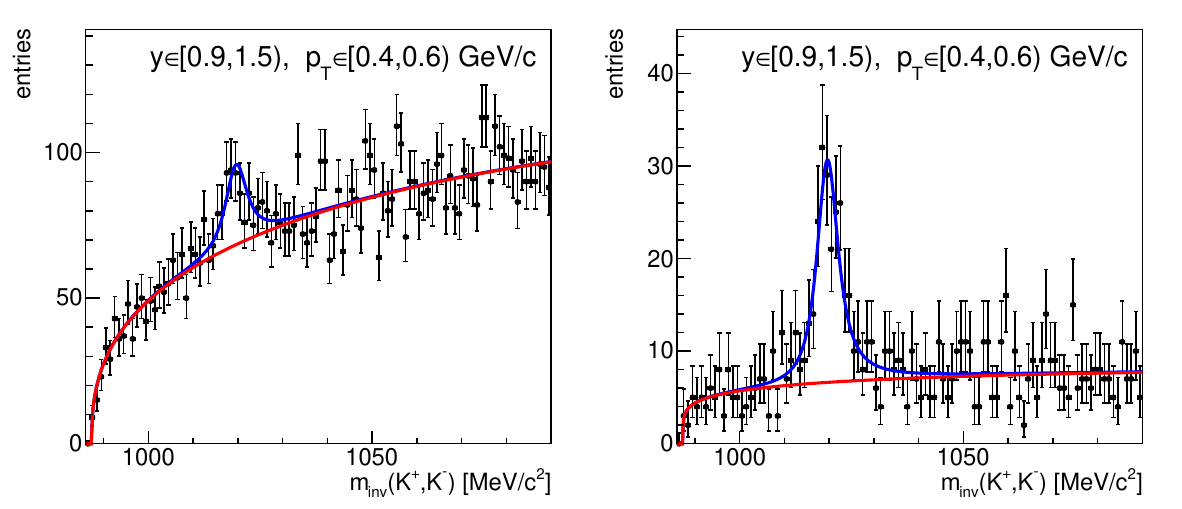}
  \caption{Example of a simultaneous tag-and-probe fit done in the final step
    of the fitting strategy to determine the raw \phi yield for one of 2D
    phase space bins of the \SI{158}{\GeVc} data. The tag (probe) sample is shown in the left (right) panel. The rapidity and \pt intervals are indicated in the figures. 
  }
  \label{fig:Example2DBinFit}
\end{figure*}
The procedure used to extract the \phi yields follows the approach introduced by the LHCb~\cite{bib:LHCb_phi_7TeV} and ATLAS~\cite{bib:ATLAS_phi_7TeV} collaborations. It is called the "tag-and-probe method" and automatically corrects for losses due to kaon candidate identification inefficiencies.
The procedure uses two data samples which differ only in the particle identification conditions. Either both partners or at least one partner of the pair are required to fulfill the PID condition selecting a kaon candidate. The former requirement leads to the probe sample of particle pairs entering the invariant mass distribution (see \figref{fig:TagAndProbeFit}). The tag sample is shown in the left panel of \figref{fig:TagAndProbeFit}.
The difference between the probe sample and the tag sample is a much better signal to background ratio in the former, because of the more complete PID information. The large increase of the background in the latter is predominantly caused by misidentified pions. This gives room for a signal from the decay of the \Kstar resonance visible as a bump above the background at
about \SI{1075}{\MeVcsq}.
\par
The simultaneous description of the invariant mass distributions built from the tag and probe samples has a new parameter~\PIDeff which is the efficiency of kaon selection (i.e.\ the probability that the kaon is accepted by the PID cut). It correlates 
the total number of \phi mesons (\Nphi) in the event ensemble with the number of \phi mesons in tag sample \Nt and in the probe sample \Np. 
For the tag sample the correlation reads:
\begin{subequations}
  \label{eq:NtNpDefs}
  \begin{align}
    \Nt\FitPars & = \Nphi \PIDeff (2 - \PIDeff) \,, \\
\intertext{while that in the probe sample is}
    \Np\FitPars & = \Nphi \PIDeff^2 \,.
  \end{align}
\end{subequations}
\par
\par
The function used to fit simultaneously both the tag and the probe invariant mass distributions reads:
\begin{equation}
  \label{eq:TagAndProbeFitFunc}
  f(\m) =
  \begin{cases}
    \IfEPJC{
      \begin{split}
        \Nt\FitPars V(\m; \mphi, \sigma) \\ + \NbkgT \Bt(\m; \ARGUSkt, \ARGUSpt)
      \end{split}
    }{
      \Nt\FitPars V(\m; \mphi, \sigma) + \NbkgT \Bt(\m; \ARGUSkt, \ARGUSpt)
    }
      & \text{for the tag} \\
    \IfEPJC{
      \begin{split}
        \Np\FitPars V(\m; \mphi, \sigma) \\ + \NbkgP \Bp(\m; \ARGUSkp, \ARGUSpp)
      \end{split}
    }{
      \Np\FitPars V(\m; \mphi, \sigma) + \NbkgP \Bp(\m; \ARGUSkp, \ARGUSpp)
    }
      & \text{for the probe}
  \end{cases}
  \,,
\end{equation}
where the quantities $V(\m; \mphi, \sigma)$ are given by \equref{eq:SignalParametrization},
while $\Bt(\m; \ARGUSkt, \ARGUSpt)$ and $\Bp(\m; \ARGUSkp, \ARGUSpp)$ are the ARGUS functions (\equref{eq:argus}) describing the backgrounds for the tag and probe samples, respectively. All three expressions are normalised in such a way that the terms \Nt and \Np defined by \equref{eq:NtNpDefs} give the numbers
of signal pairs in the tag and probe spectra, while \NbkgT and
\NbkgP give the numbers of background pairs in the respective histograms. In total there are ten free parameters to be fitted to the data, four for the signal (\Nphi, \PIDeff, \mphi, $\sigma$) and six for the background (\NbkgT, \ARGUSkt, \ARGUSpt, \NbkgP, \ARGUSkp, \ARGUSpp).
\par
Note that \Nphi should be understood as the number of \phi mesons, the
daughters of which pass all track cuts apart from the PID cut. This means that this number is still subject to corrections for various effects other than PID (like e.g.\ geometrical acceptance, reconstruction as well as trigger efficiency). 

\subsection{Fitting strategy}\label{s:fitStrategy}
Due to limited statistics not all parameters of \equref{eq:TagAndProbeFitFunc}
discussed above can be fitted in each analysis bin separately. A three-step
fitting strategy was developed instead. All the fits are extended binned
log-likelihood fits (see e.g.\ \recite{bib:CowanStats}).
\par
In a first step precise values of signal shape
parameters \mphi and $\sigma$ are determined on a high statistics histogram which uses a large part of the covered phase space. The corresponding invariant mass distribution is shown in \figref{fig:SingleSpectrumFit} together with the function defined by \equref{eq:SingleFitFunc}. 
The resulting values of \mphi and $\sigma$ are fixed in further steps.
\par
In a second step the values of the PID efficiency parameter \PIDeff are determined for use in step three. In this single differential analysis five bins in rapidity are used with an integration over a broad range in transverse momentum (\figref{fig:BinningTypes:1Dy}).
In each bin of rapidity, a simultaneous tag-and-probe fit is performed using the function \equref{eq:TagAndProbeFitFunc}, with fixed signal shape parameters \mphi and $\sigma$. The resulting \PIDeff values vary from $0.61\pm~0.06$ at low rapidities to $0.93\pm~0.06$ at high rapidities.
This procedure assumes that the kaon identification efficiency does not change significantly with \pt which has been demonstrated in \recite{bib:MarcinekPhD}.
\par
Finally, in the third step of the strategy, simultaneous tag-and-probe fits are done in all selected rapidity and transverse momentum bins and provide the raw \phi yields of the one-dimensional and two-dimensional analyses.
Again the function \equref{eq:TagAndProbeFitFunc} is employed,
with fixed signal shape parameters \mphi and $\sigma$, and with \PIDeff
determined as explained above. An example is shown in \figref{fig:Example2DBinFit}.

\begin{figure*}[t!]
  \centering
  \includegraphics[width=\textwidth]{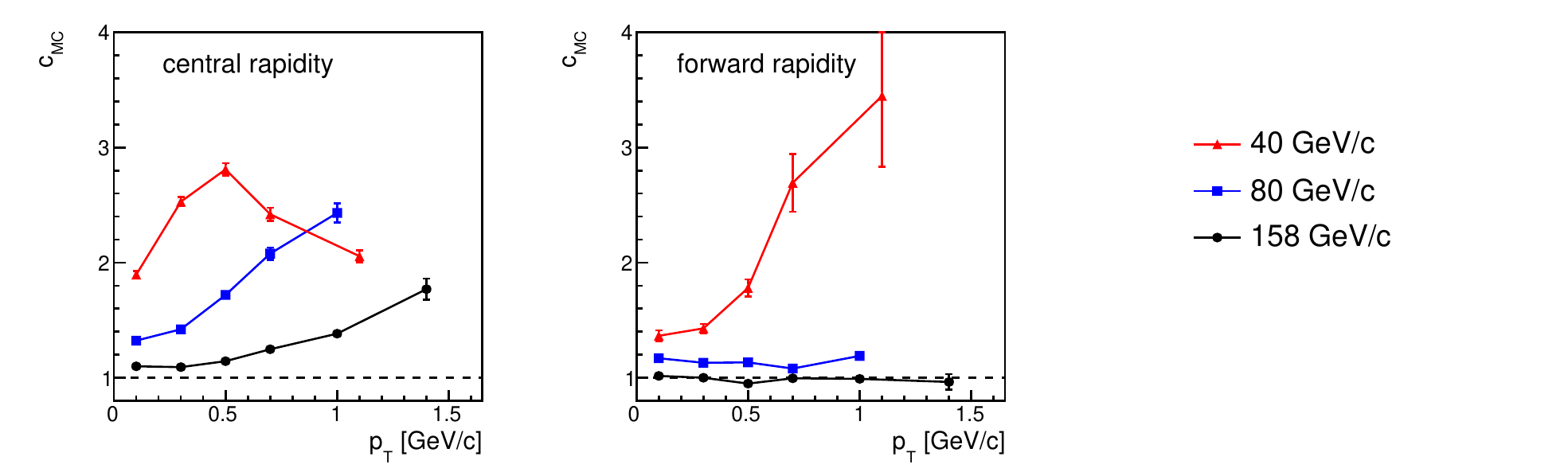}
  \caption{The Monte Carlo correction coefficient \mcCorr drawn as a function
    of transverse momentum for one central and one forward rapidity bin, at
    beam momenta of \SI{40}{\GeVc} (red), \SI{80}{\GeVc} (blue) and \SI{158}{\GeVc} (black). The central rapidity bin (left) is always (0.0,0.3) while the forward rapidity bin (right) is (0.9,1.2) for \SI{40}{\GeVc} beam momentum and (0.9,1.5) at other beam momenta.
  }
  \label{fig:158Breakdown}
\end{figure*}
\subsection{Corrections}\label{s:corr}
The present analysis includes corrections for the branching ratio of the \phi decay into $\Kp$ $\Km$ and the cut-off used in the integration of the resonance signal. A Monte Carlo-based procedure provides the corrections for losses due to the vertex cuts, geometrical acceptance of kaons coming from \phi decays, the track reconstruction inefficiency including bin migration due to momentum resolution, and the event losses introduced by the minimum bias trigger.
\par
The fully corrected double differential spectrum of the number of \phi mesons per event is given by
\begin{equation}
  \label{eq:DoubleDiffSpectrum}
  \doubleDiffSpectrum =
    \frac{\Nphi}{\Nev \DD\pt \DD\y} \times
    \frac{\intCorr \mcCorr}{\branching} \,,
\end{equation}
where the first term is the normalized raw spectrum obtained in the analysis with the bin widths $\DD\pt$ and $\DD\y$;
\intCorr~is the correction due to the integration cut-off and of order of
6\%. The \branching is taken from \recite{bib:PDG}.
\par
The Monte Carlo correction factors \mcCorr in various \yptBin bins are shown in Fig.~\ref{fig:158Breakdown} as a function of the transverse momentum for one central and one forward bin in rapidity at three collision energies. The correction clearly depends on both \y and \pt, and also on collision energy. The latter is not surprising as different beam energies mean different boosts of the emitted particles which cause different opening angles and thus increasing acceptance losses with decreasing energy. The correction coefficient can be both above and below unity. The latter is caused by trigger and vertex cut losses which both tend to eliminate low multiplicity \pp events and to artificially enhance the measured \phi yield. 
A complete description of the correction procedures with their uncertainties can be found 
in \recite{bib:MarcinekPhD}. The systematic errors are addressed in the next paragraph.
\begin{figure*}
  \centering
  \subfloat[\SI{158}{\GeVc}]{
    \label{fig:Uncertainties:158}
    \includegraphics[height=0.45\textheight]{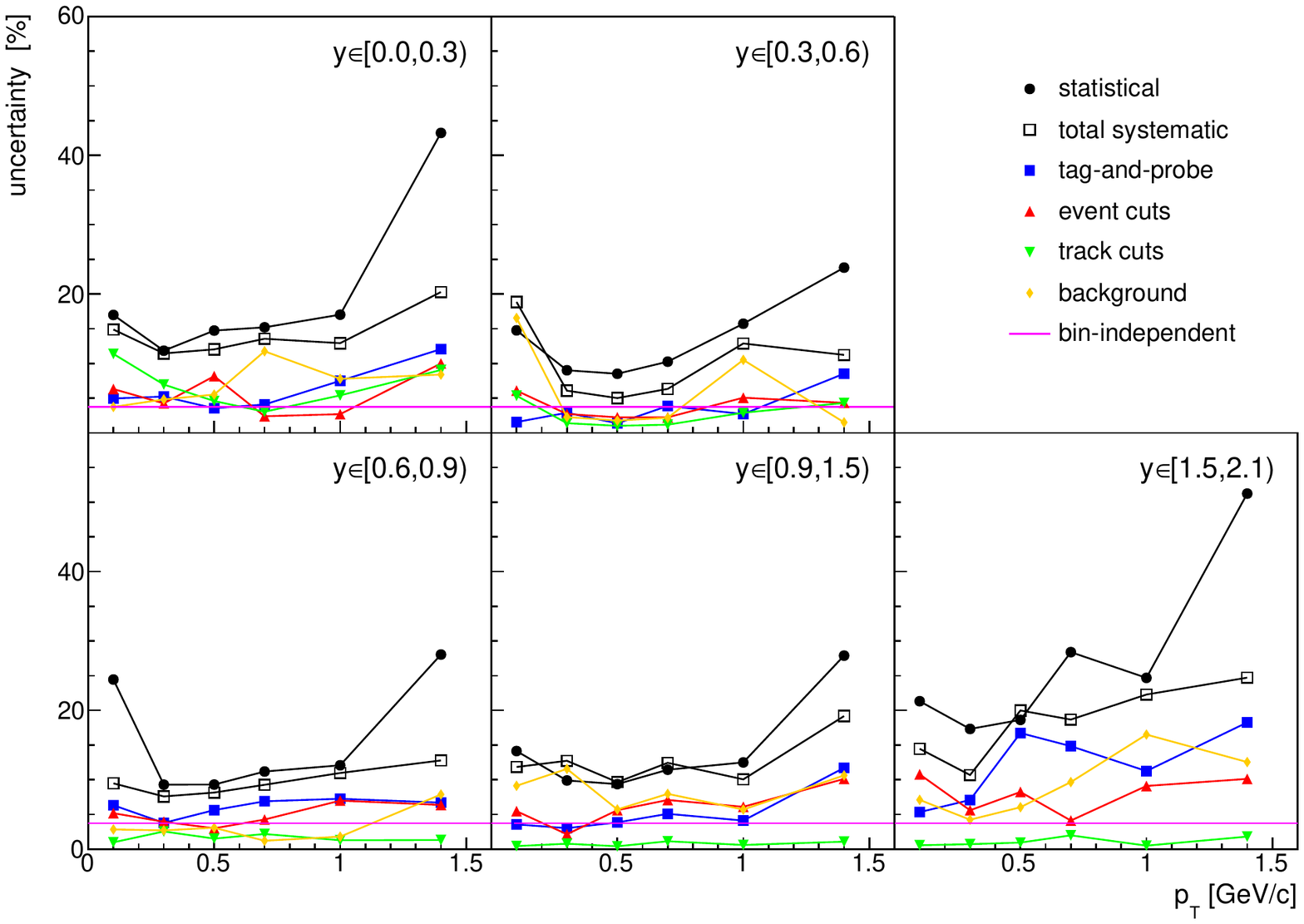}
  }
  \\
  \subfloat[\SI{80}{\GeVc}]{
    \label{fig:Uncertainties:80}
    \includegraphics[height=0.45\textheight]{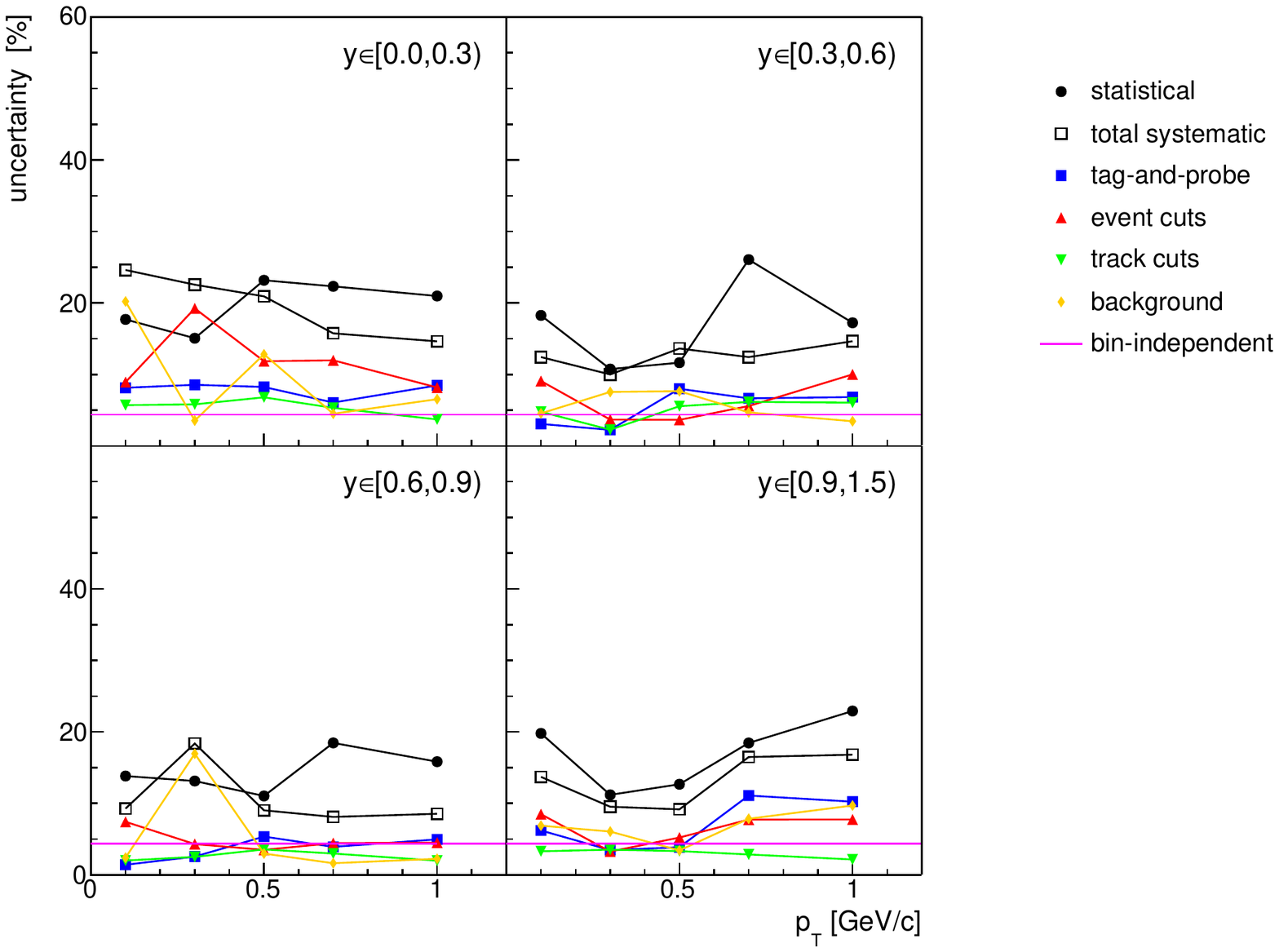}
  }
  \caption{Comparison of statistical and systematic uncertainties for the
    double differential analysis of \SI{158}{\GeVc} (a) and \SI{80}{\GeVc} data (b).
    The \pt dependences are shown for different rapidity intervals.
    Total systematic uncertainty is calculated by adding contributions in
    quadrature.
  }
  \label{fig:Uncertainties}
\end{figure*}
\begin{figure*}
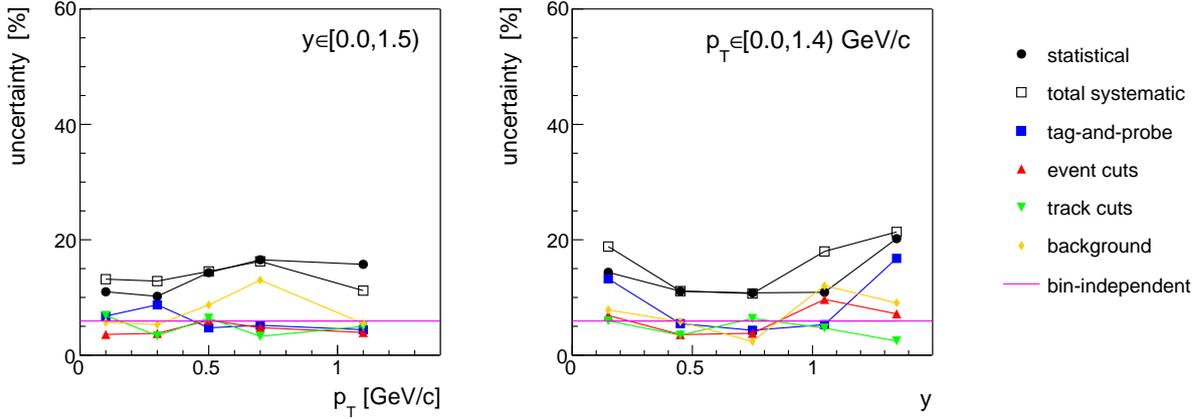

  \centering
  \resultsPtY{1}{pp40_y_pt_err.pdf}{pp40_pt_y_err.pdf}
  \caption{Comparison of statistical and systematic uncertainties for the
    single differential analysis of \SI{40}{\GeVc} data. 
    The \pt (left) and rapidity (right) dependences are shown for the indicated rapidity and \pt intervals. The total systematic uncertainties are calculated by adding the five contributions in quadrature.
  }
  \label{fig:Uncertainties:40}
\end{figure*}
\par
\begin{table}
  \centering
  \caption{Bin-independent systematic uncertainties. \enquote{Total} is calculated by
    adding the contributions in quadrature.
  }
  \begin{tabular}{@{}lc@{\qquad}c@{\qquad}c@{}}
    \toprule
      & \multicolumn{3}{c}{uncertainty value [\%]} \\
    \cmidrule(l){2-4}
    Source & \SI{158}{\GeVc} & \SI{80}{\GeVc} & \SI{40}{\GeVc} \\
    \midrule
    branching ratio & 1 & 1 & 1 \\
    fitting constraints & 2 & 3 & 4 \\
    \phi signal & 3 & 3 & 3 \\
    correction averaging & --- & --- & 3 \\
    \midrule
    Total & 6 & 7 & 8 \\
    \bottomrule
  \end{tabular}
  \label{tab:uncertainties}
\end{table}
The choice of the integration range used to obtain the \phi yield from the signal parametrization curve has a negligible effect on the magnitude of the respective correction factor (1.06).
Similarly, variations of the \phi production model used in the Monte Carlo correction 
averaging in case of the single differential analysis
does not change the results significantly. Bin-independent systematic uncertainties arise from the choice of the fitting constraints, the \phi signal parametrization, and the correction averaging. They are listed in \tabref{tab:uncertainties}. 
The particle identification efficiencies which are determined by the tag and probe analysis may not be constant in the considered rapidity and transverse momentum bins. The resulting systematic uncertainties are due to
shortcomings in the particle identification procedures which may generate systematic errors of the tag and probe analysis. The corresponding uncertainties can be read off the diagrams presented in \figref{fig:Uncertainties,fig:Uncertainties:40}. They stay well below the statistical errors which are added for comparison. Also shown are the systematic errors introduced by the event and track cuts which may occur, if the generated MC events do not precisely enough reproduce the experimental distributions of the cut variables. A further source of systematic uncertainty is the choice of the background function for the fit of the invariant mass distribution. The \phi mass is near to the 
two-kaon mass threshold. The background at threshold may have (small) contributions of
correlated kaons from \fzero or \azero decays. Also at about \SI{1075}{\MeVcsq}
possible correlated pairs of kaons and misidentified pions from the decay of
the \Kstar resonance may appear, especially in the tag sample. To estimate the associated systematic uncertainty the fit range was varied and the resulting yield differences were used as one set of systematic errors. A second set was obtained by replacing the ARGUS function by a function consisting of  templates of the combinatorial background (from event mixing), of \Kstar resonance decays, and of \fzero- or \azero-like decays. The largest of the two estimates was taken bin-by-bin as the systematic error.
The total systematic error is calculated by adding all contributions in quadrature and stays always below or close to the statistical error in \figref{fig:Uncertainties,fig:Uncertainties:40}.

\par

\section{Results}\label{s:Results_paper}

Yields of \phi mesons have been determined as function of transverse momentum (up to 6 bins) and rapidity (up to 5 bins) in \pp interactions at beam momenta of \SI{158}{\GeVc} and \SI{80}{\GeVc}. These are the first double differential measurements of \phi production in proton-proton collisions at CERN SPS energies. Due to low statistics \phi yields at \SI{40}{\GeVc} have only been obtained as function of transverse momentum (5 bins) and rapidity (5 bins) (integrated over rapidity and transverse momentum, respectively).

\begin{figure*}
  \centering
  \subfloat[\SI{158}{\GeVc}]{
    \label{fig:PtResults:158}
    \includegraphics[height=0.45\textheight,page=2]{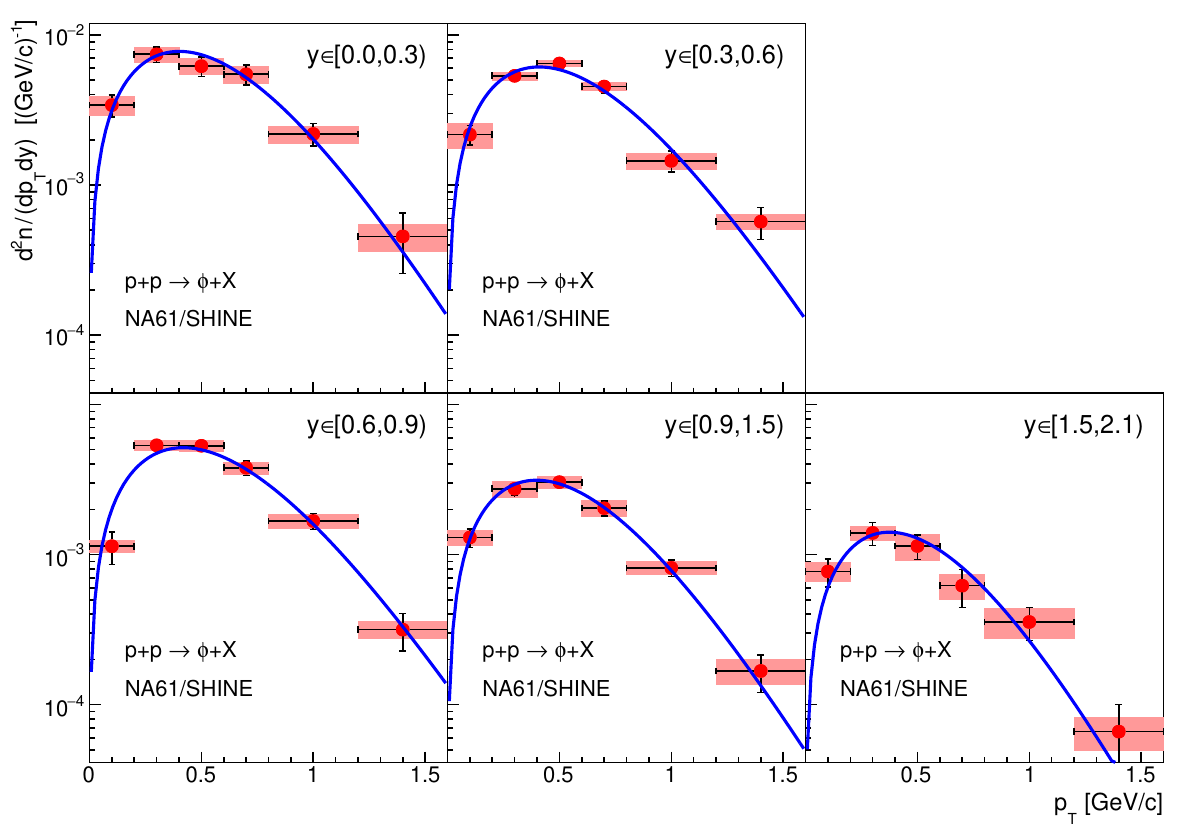}
  }
  \\
  \subfloat[\SI{80}{\GeVc}]{
    \label{fig:PtResults:80}
    \includegraphics[height=0.45\textheight,page=2]{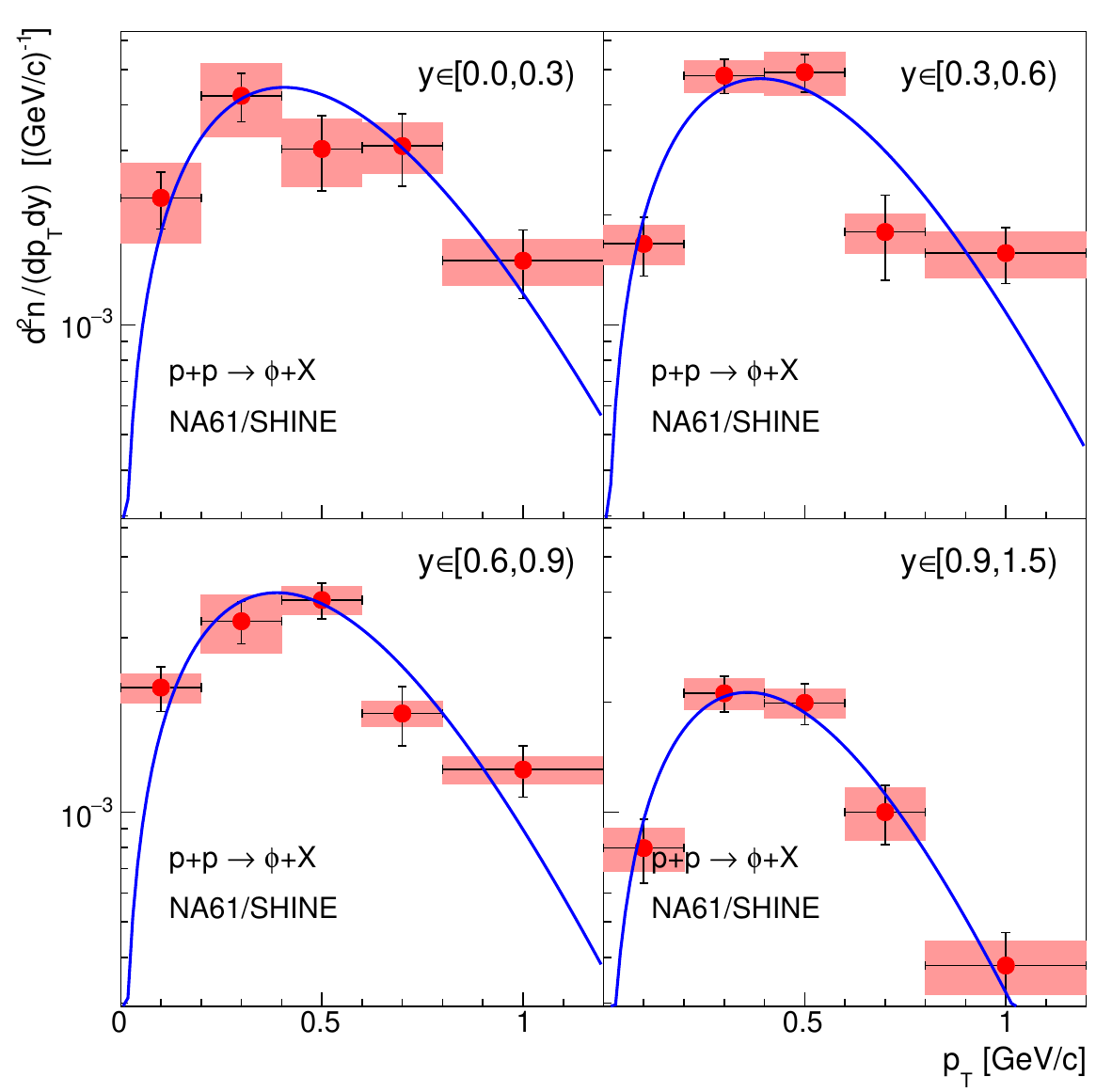}
  }
  \caption{Transverse momentum spectra in 6 rapidity bins for \SI{158}{\GeVc} and 5 rapidity bins for
    \SI{80}{\GeVc} data with statistical errors (vertical bars) and systematic errors (red shaded bands).
    The horizontal bars indicate the bin size.
    Curves are fits of function \equref{eq:NormalizedThermal}.
  }
  \label{fig:PtResults}
\end{figure*}
\begin{figure*}
  \centering
  \includegraphics[width=0.5\textwidth,page=2]{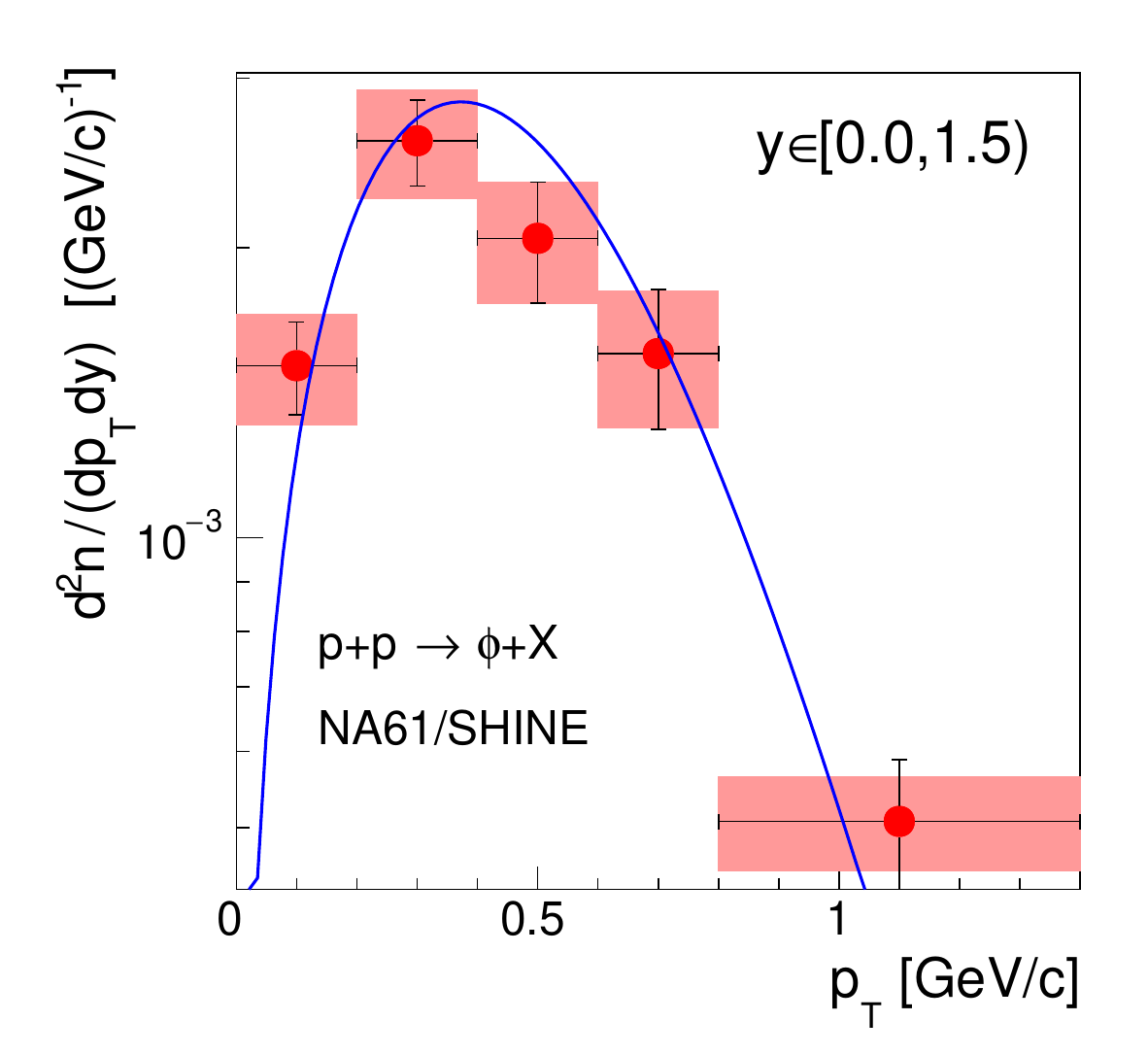}
  \caption{Transverse momentum spectrum integrated over rapidity for \SI{40}{\GeVc} data with statistical errors (vertical bars) and systematic errors (red shaded bands). 
  The horizontal bars indicate the bin size.
  Solid blue curve is a fit of the function defined in \equref{eq:NormalizedThermal}. 
  }
  \label{fig:PtResults:40}
\end{figure*}
\begin{figure*}
  \centering
  \stdMoveLeft
  \subfloat[\SI{158}{\GeVc}]{
    \label{fig:mtResults:158}
    \includegraphics[width=0.5\textwidth]{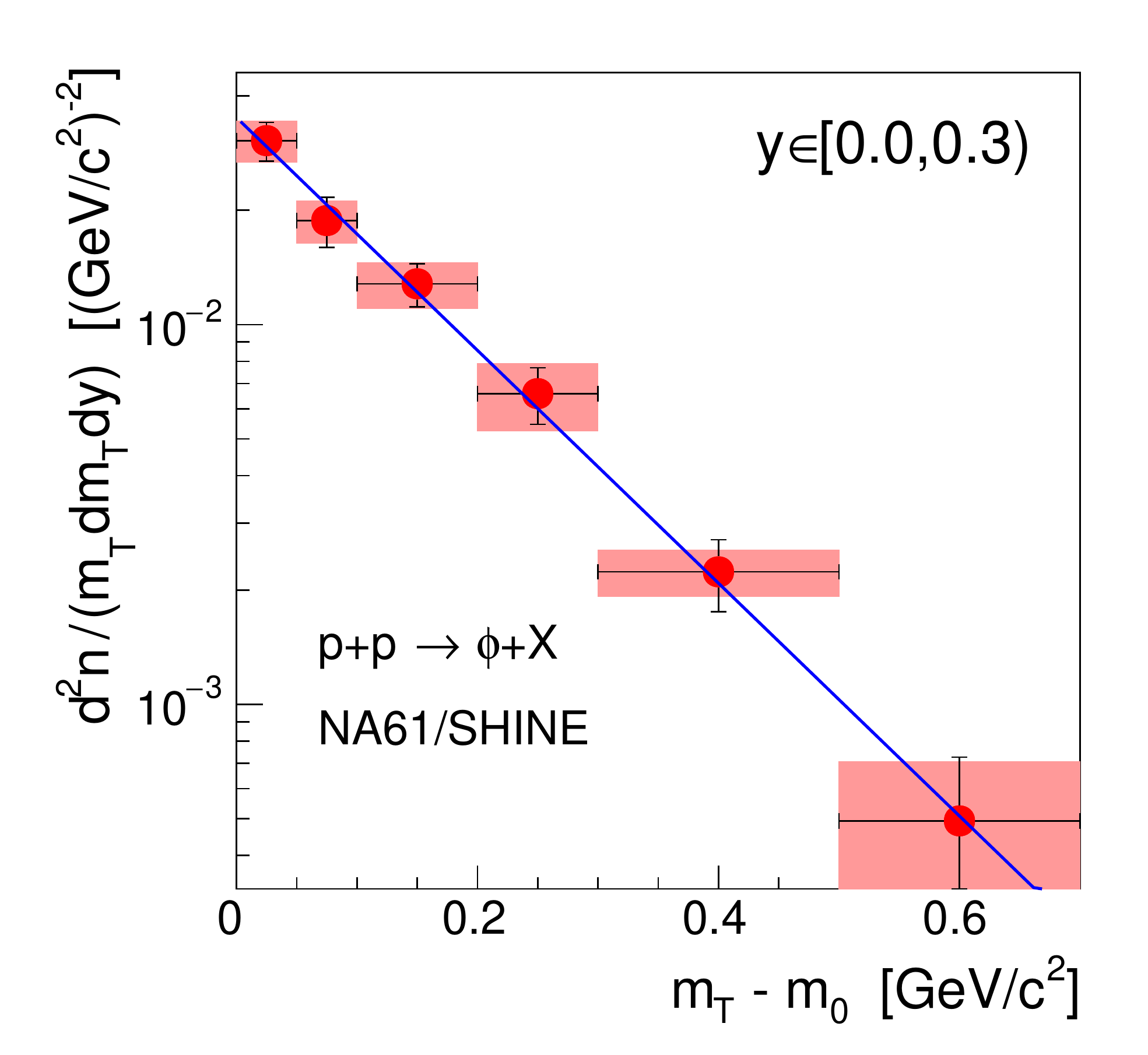}
  }
  \subfloat[\SI{80}{\GeVc}]{
    \label{fig:mtResults:80}
    \includegraphics[width=0.5\textwidth]{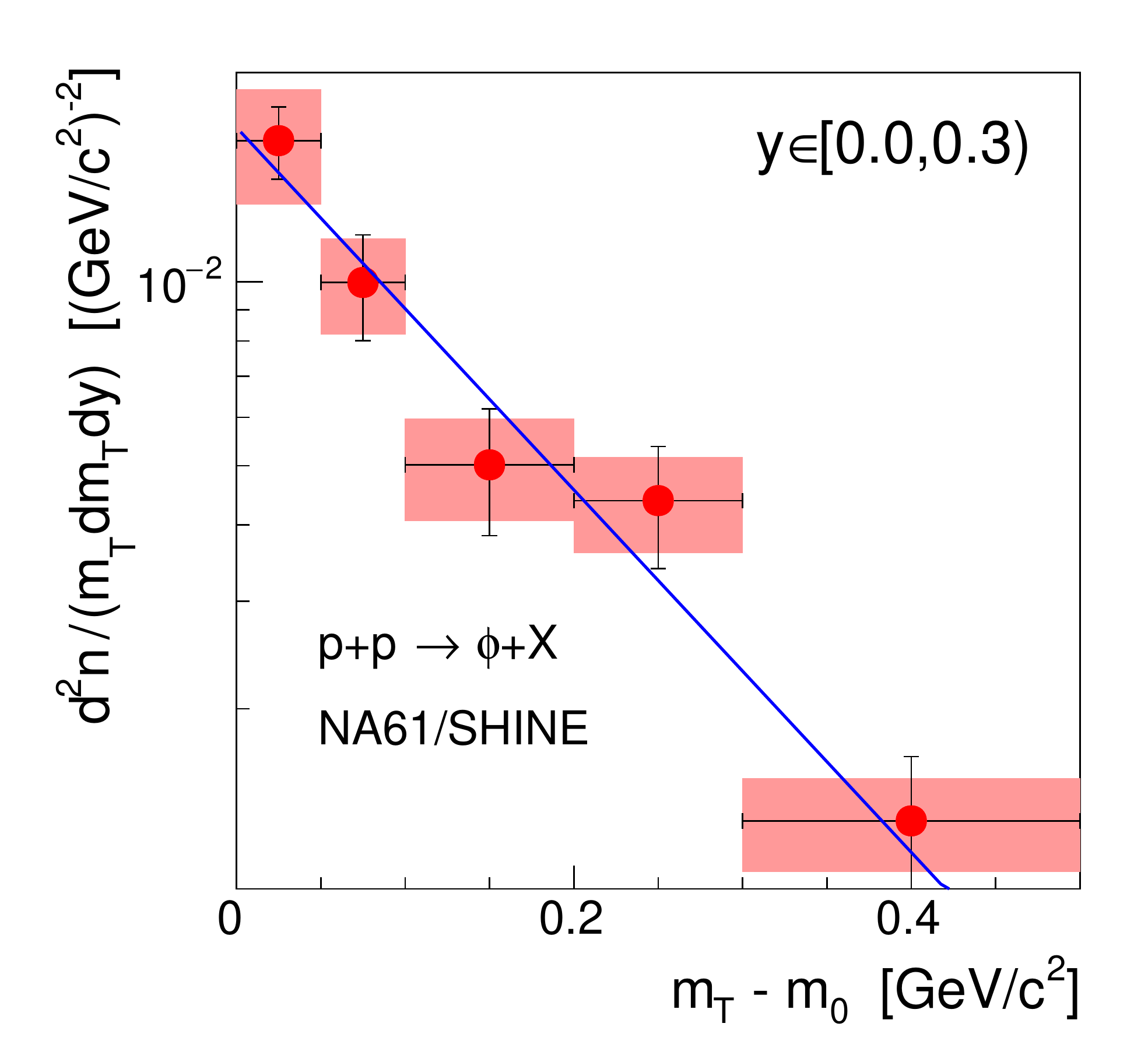}
  }
  \caption{Transverse mass spectra at midrapidity for \SI{158}{\GeVc} (left) and \SI{80}{\GeVc} (right). The straight lines are the fits of \equref{eq:NormalizedThermal} to the data points.
}
  \label{fig:mtResults}
\end{figure*}
\begin{figure*}[!bt]
  \centering
  \stdMoveLeft
  \subfloat[\SI{158}{\GeVc}]{
    \label{fig:SpectralPars:158}
    \includegraphics[width=\textwidth,page=2]{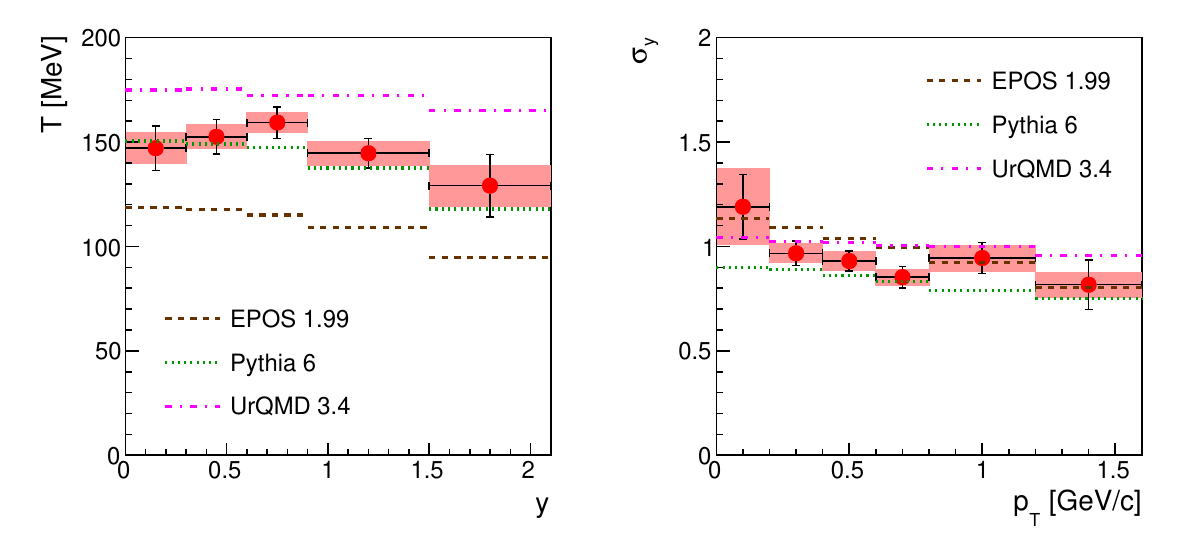}
  }
  \\
  \stdMoveLeft
  \subfloat[\SI{80}{\GeVc}]{
    \label{fig:SpectralPars:80}
    \includegraphics[width=\textwidth,page=2]{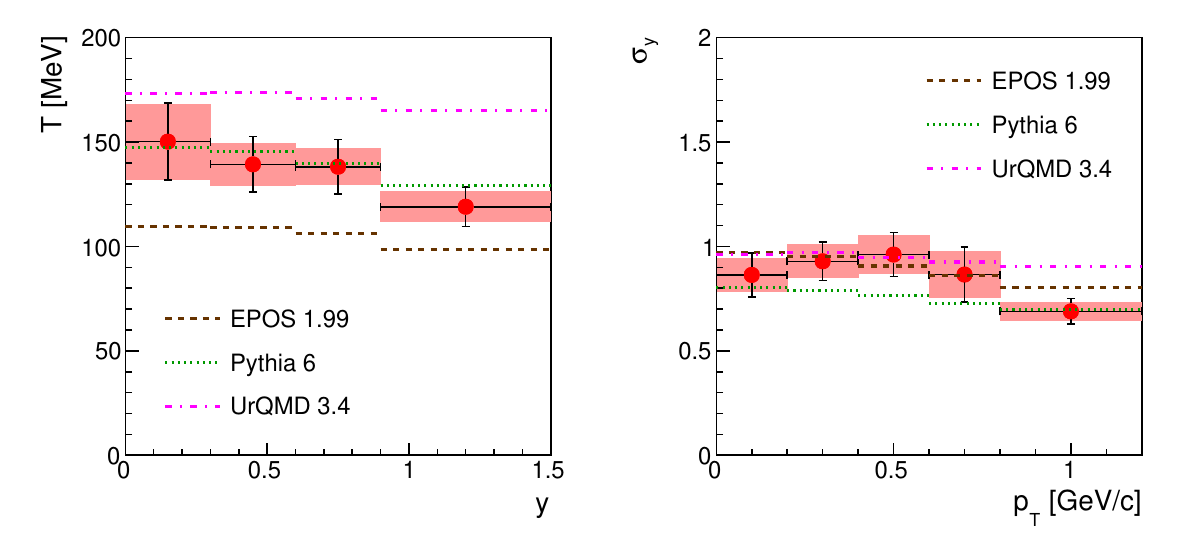}
  }
  \caption{Dependence of the slope parameter $T$ on rapidity (left) and the width $\sigma_y$ of the rapidity distributions on \pt (right) for \SI{158}{\GeVc} and \SI{80}{\GeVc} data 
  with statistical errors (vertical bars) and systematic errors (red bands).
    The horizontal bars indicate the bin size.
}
  \label{fig:SpectralPars}
\end{figure*}
The resulting transverse momentum spectra are shown in \figref{fig:PtResults:158} for the \SI{158}{\GeVc}, in \figref{fig:PtResults:80} for the \SI{80}{\GeVc} and \figref{fig:PtResults:40} for the \SI{40}{\GeVc} data. If the yields, divided by the transverse mass (\mt), are plotted as function of \mtmRest instead exponential shapes emerge as shown for the midrapidity bins in \figref{fig:mtResults:158} and \figref{fig:mtResults:80}. This suggests to fit the transverse momenta (mass) spectra with the function defined in \equref{eq:NormalizedThermal} 
\begin{equation}
  \label{eq:NormalizedThermal}
  \fNormTherm = A \times \pt \exp(-\frac{\mt}{\temperature}) \,,
\end{equation}
to characterize the shape of the spectra by a single slope parameter \temperature and to estimate the yield outside of the acceptance (at high transverse momenta).  For \SI{158}{\GeVc} and \SI{40}{\GeVc} these contributions are below \SI{1}{\percent} for all rapidity bins, while for \SI{80}{\GeVc} they are of the order of \SIrange{1}{4}{\percent}. The function \fNormTherm describes the experimental data within errors in all rapidity bins.
The rapidity dependence of the slope parameter \temperature, often called effective temperature, is given in \figref{fig:SpectralPars:158} (left) for the \SI{158}{\GeVc} and
in \figref{fig:SpectralPars:80} (left) for the \SI{80}{\GeVc} data.
\begin{figure*}[!btp]
  \centering
  \stdMoveLeft
  \subfloat[\SI{158}{\GeVc}]{
    \label{fig:Rapidity:158}
     \includegraphics[width=0.33\textwidth,page=2]{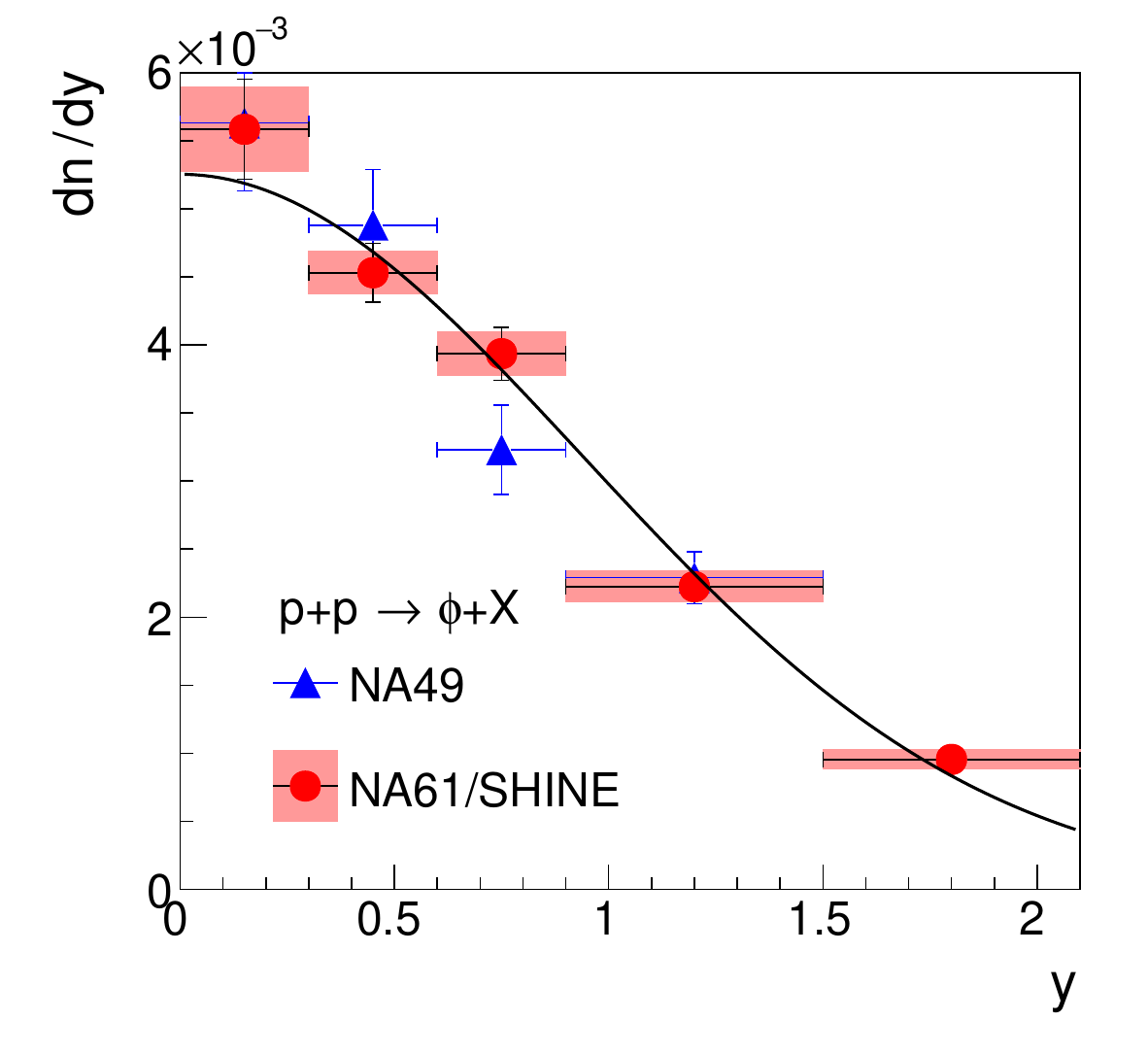}
 }
  \subfloat[\SI{80}{\GeVc}]{
    \label{fig:Rapidity:80}
    \includegraphics[width=0.33\textwidth,page=2]{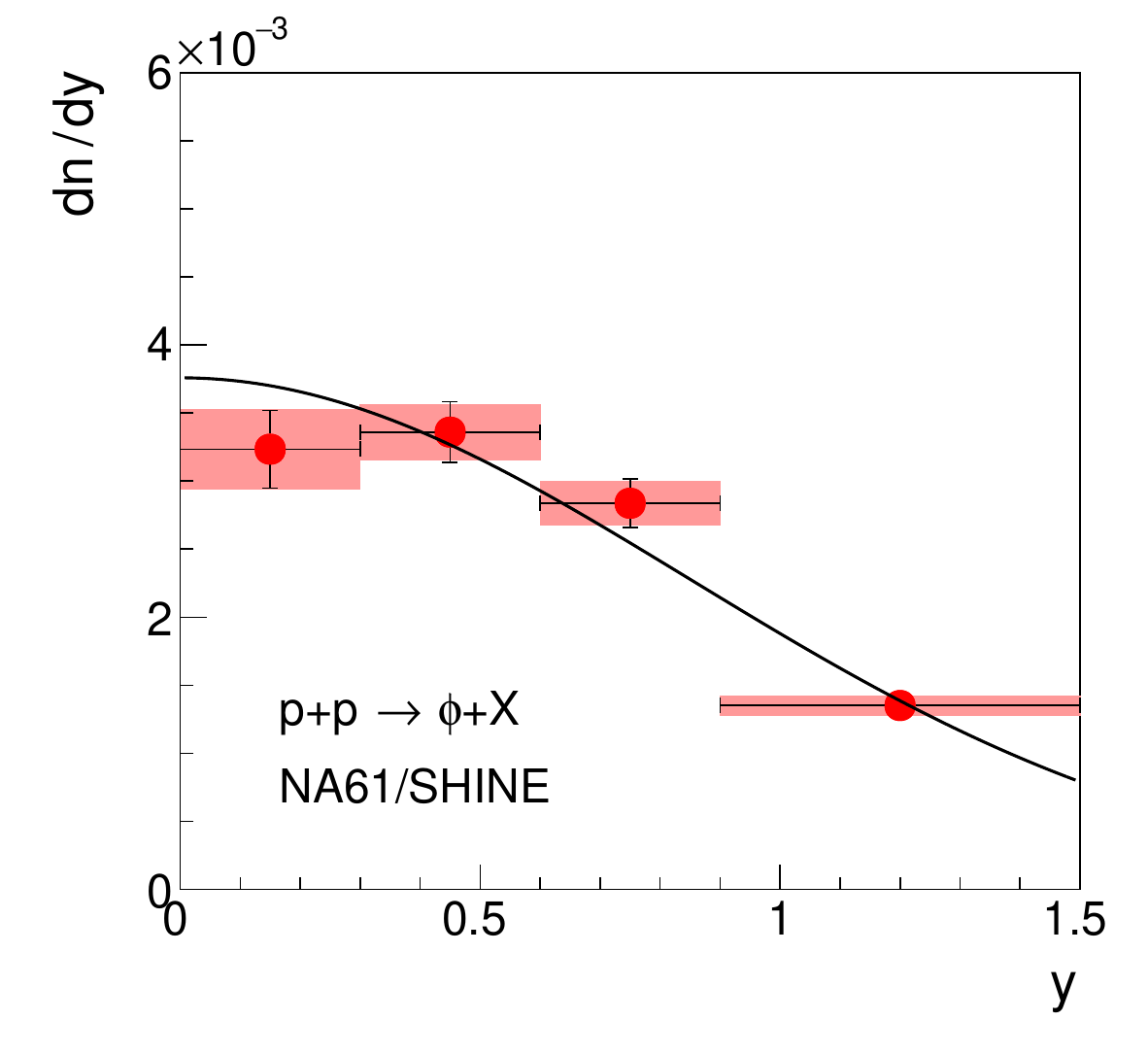}
}
  \subfloat[\SI{40}{\GeVc}]{
    \label{fig:Rapidity:40}
    \includegraphics[width=0.33\textwidth,page=2]{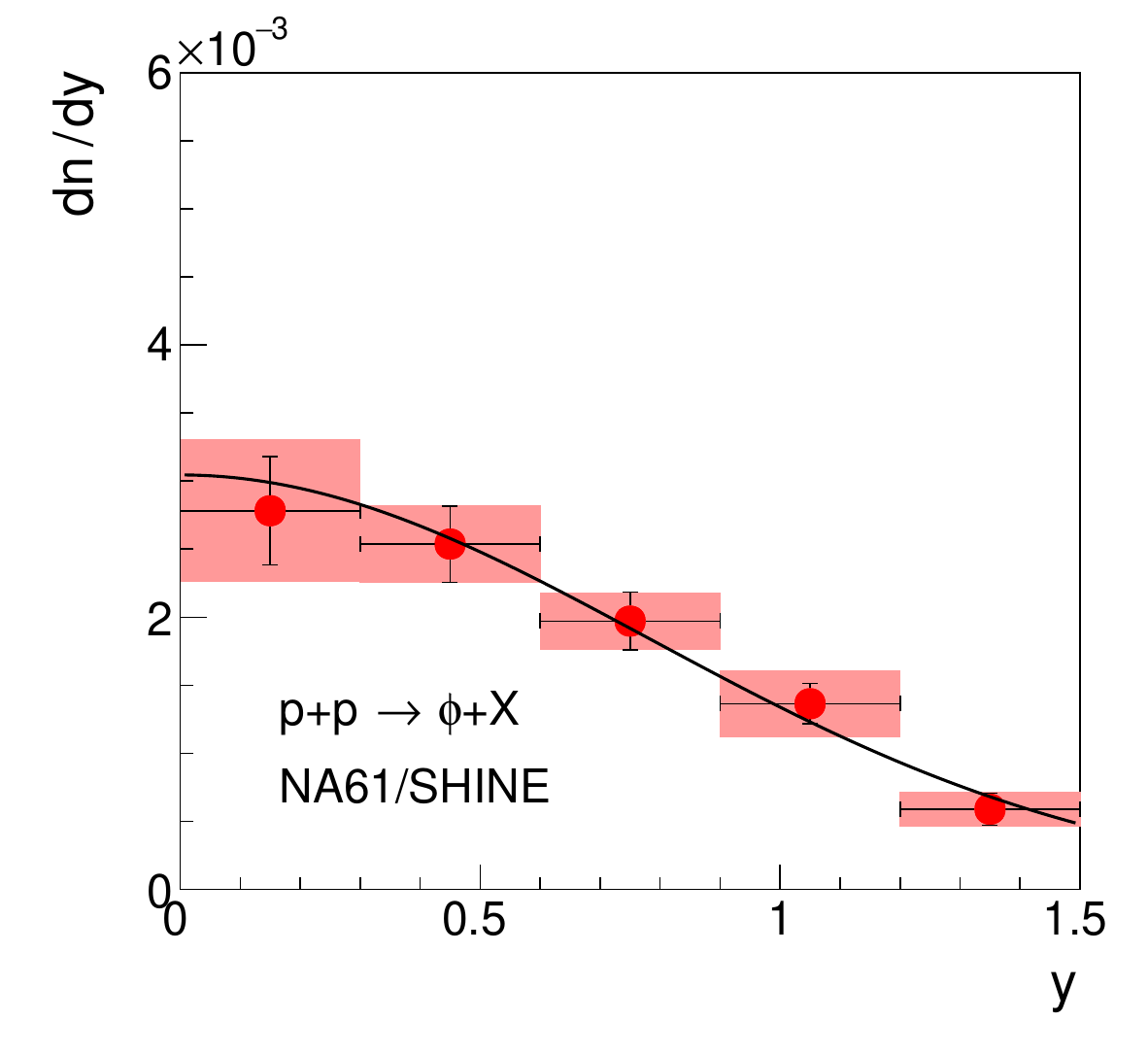}
}
\caption{Rapidity spectra for \SI{158}{\GeVc}, \SI{80}{\GeVc}, and \SI{40}{\GeVc} data with statistical errors 
    (vertical bars) and systematic errors (red bands). The horizontal bars indicate the bin size.
    NA49 points (triangles) come from \recite{bib:NA49phi2000}. Solid curves are Gaussian fits
    (\equref{eq:NormalizedGaus}) 
  }
  \label{fig:Rapidity}
\end{figure*}
Rapidity yields are obtained by summing the content of the corresponding \pt spectra and adding the corrections for the unmeasured regions. The resulting rapidity spectra (in the centre-of-mass) are shown in \figref{fig:Rapidity:158} (\SI{158}{\GeVc}), \figref{fig:Rapidity:80} (\SI{80}{\GeVc}), and \figref{fig:Rapidity:40} (\SI{40}{\GeVc}) in the forward hemisphere.
Their shapes can be approximated by Gaussian distributions. The corresponding fits with
\begin{equation}
  \label{eq:NormalizedGaus}
  \fNormGaus =
     A \times \exp(-\frac{\y^2}{2 \sigmaY^2}) \,,
\end{equation}
provide width parameters $\sigma_y$ for each \pt bin which are shown in \figref{fig:SpectralPars}~(right).
\par
\begin{table*}[!btp]
  \centering
  \caption{Parameters deduced from rapidity distributions for all analysed beam
    momenta. The first uncertainty is statistical, the second one systematic.
  }
  \label{tab:RapidityFits}
  \begin{tabular}{@{}lcccc@{}}
    \toprule
    \pbeam [\si{\GeVc}] & \sigmaY & \totalYield [$10^{-3}$] & \midrapYield [$10^{-3}$] & $\chi^2 / \text{ndf} $ \\
    \midrule
    158 & \ensuremath{0.938 \pm 0.027 \pm 0.023
}\xspace & \ensuremath{12.56 \pm 0.33 \pm 0.32
}\xspace & \ensuremath{5.25 \pm 0.19 \pm 0.15
}\xspace & 0.94 \\
    80 & \ensuremath{0.850 \pm 0.040 \pm 0.033
}\xspace & \phantom{1}\ensuremath{7.89 \pm 0.29 \pm 0.39
}\xspace & \ensuremath{3.76 \pm 0.20 \pm 0.19
}\xspace & 1.73 \\
    40 & \ensuremath{0.780 \pm 0.047 \pm 0.053
}\xspace & \phantom{1}\ensuremath{5.87 \pm 0.35 \pm 0.44
}\xspace & \ensuremath{3.05 \pm 0.25 \pm 0.28
}\xspace & 0.27 \\
    \bottomrule
  \end{tabular}
\end{table*}

\IfEPJC{
  \begin{figure*}[bp!]
}{
  \begin{figure*}[bt!]
}
  \centering
  \stdMoveLeft
  \subfloat[]{
    \label{fig:YieldRatios:phi}
    \includegraphics[width=0.5\textwidth,page=1]{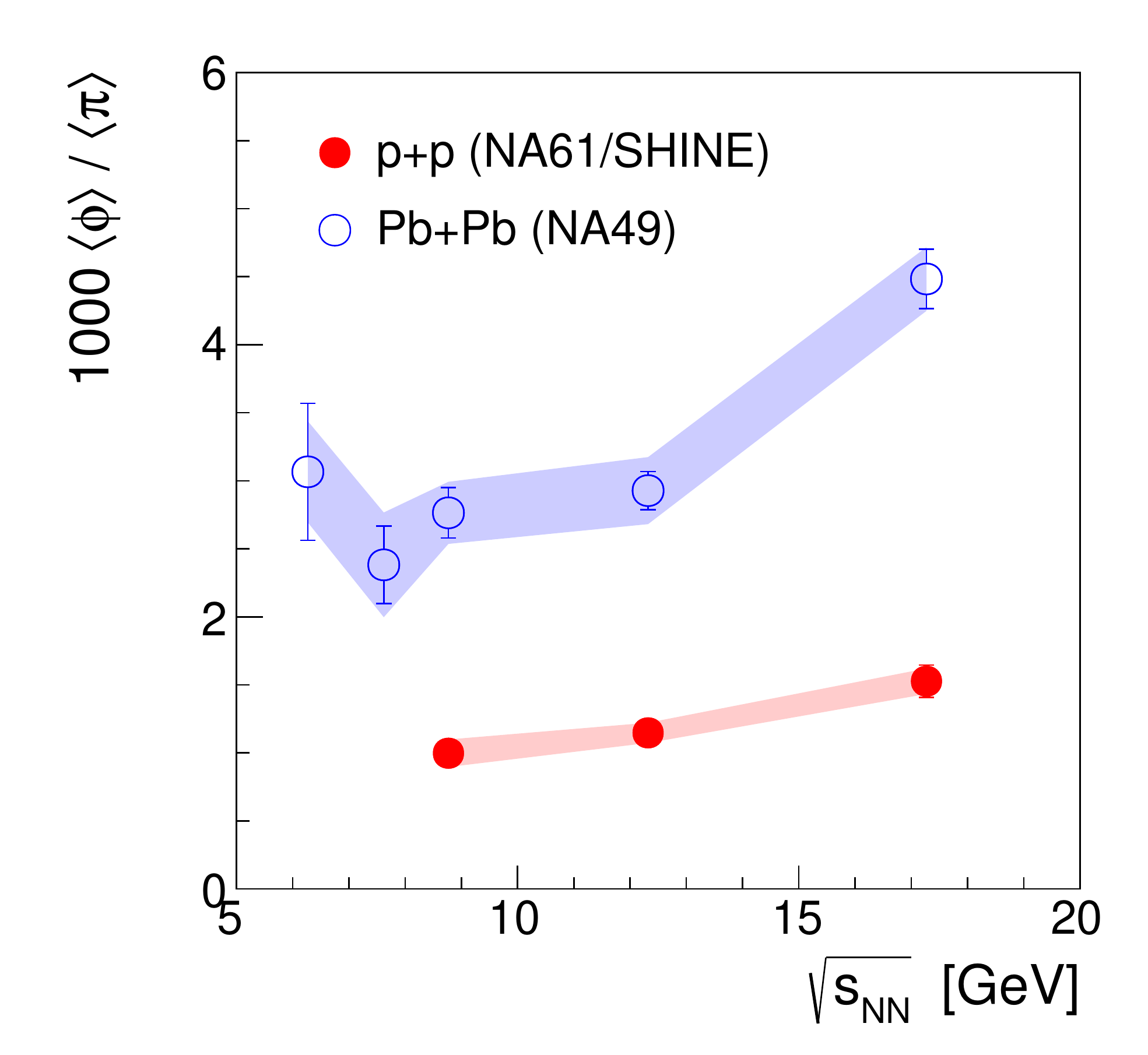}
  }
  \subfloat[]{
    \label{fig:YieldRatios:double}
    \includegraphics[width=0.5\textwidth,page=2]{yield_ratios.pdf}
  }
  \caption{Energy dependence of \subref{fig:YieldRatios:phi} ratios of total
    yields of \phi mesons to mean total yields for pions (\equref{eq:piYield})
    in \pp and \PbPb, \subref{fig:YieldRatios:double} double ratios (see text).
    Full red circles correspond to results of this analysis, \PbPb data come
    from NA49~\cite{bib:NA49phi2008, bib:NA49Onset2002, bib:NA49Onset2008},
    while \pp kaon and pion data are taken from \recite{bib:PulawskiPhD}.
    Possible correlations of uncertainties of yields within the same reaction
    are neglected. This may lead to a slight overestimation of the indicated
    uncertainties.
  }
  \label{fig:YieldRatios}
\end{figure*}
Total \phi yields (\totalYield) are obtained by summing the content of the rapidity spectra and adding a correction for the extrapolation into the unmeasured beam and target rapidity regions, which is obtained from the Gaussian fits. The unmeasured tail contributions to \totalYield are about \SI{3}{\percent} for
\SI{158}{\GeVc}, \SI{7}{\percent} for \SI{80}{\GeVc}, and \SI{5}{\percent} for \SI{40}{\GeVc}. 
The results for \totalYield, the width parameters $\sigma_y$ and the midrapidity yield \midrapYield  are listed in \tabref{tab:RapidityFits}.
\par
The \phi multiplicity at \SI{158}{\GeVc} reported here ($(\resultOnlyStat{pp158_total_yield}) \times 10^{-3}$) is in good agreement with the result quoted in \recite{bib:NA49phi2000} (\num{12 \pm 1.5 E-3}). The latter is more than two times less accurate, mainly because of smaller rapidity coverage and the resulting large uncertainty of the extrapolation in rapidity. The \NASixtyOne result for the \sigmaY~parameter (\resultOnlyStat{pp158_sigma_y}) also agrees with the NA49 finding of \num{0.89 \pm 0.06} \cite{bib:NA49phi2000} within quoted uncertainties. In the present analysis the slope parameter \temperature of the transverse momentum distribution was determined in the same phase bin as used by the NA49 collaboration and found at \SI{146 \pm 5}{\GeV} within errors compatible with the NA49 measurement of \SI{169 \pm 17}{\GeV}.
\par
The \NASixtyOne results are now compared to three microscopic models. These are
\Epos 1.99~\cite{bib:EPOS2006, bib:EPOS2009} and \Pythia 6.4.28~\cite{bib:PythiaManual} from the \CRMC 1.6.0
package~\cite{bib:CRMC} and \UrQMD 3.4~\cite{bib:UrQMD1998, bib:UrQMD1999}. In \Epos the \phi width had to be adjusted to
the PDG value. In case of \Pythia, the main Perugia 2011 tune 350~\cite{bib:PerugiaTunes} is used. 
The results of the model calculations on \pt and rapidity spectra are compared to the measurements in 
\figref{fig:SpectralPars}. \Pythia reproduces the shapes of the \pt spectra quite well,
while \UrQMD produces slightly harder and \Epos slightly softer spectra. This applies to both data sets \SI{158}{\GeVc} and \SI{80}{\GeVc}. 
The widths of the rapidity distributions are reproduced by the models within the systematic errors.
\par
\NewMathSymbol{\piYield}{\expval{\pi}}
\NewMathSymbol{\yieldsRatio}{\totalYield / \piYield}
\NewMathSymbol{\yieldsRatioQty}{\pqty{\yieldsRatio}}
\Figref{fig:YieldRatios:phi} presents ratios of total yields of \phi mesons to mean total yields of pions in \pp and central \PbPb~\cite{bib:NA49phi2008}
collisions as a function of energy per nucleon pair. Mean total yields for pions are calculated as in \recite{bib:NA49phi2008} :
\begin{equation}
  \label{eq:piYield}
  \textstyle
  \piYield = \dfrac{3}{2} \pqty{\expval{\pi^+} + \expval{\pi^-}} \,.
\end{equation}
The results confirm the enhancement of \phi production (normalized to pions) in the SPS energy range. This enhancement can be quantified by the double ratio (see \figref{fig:YieldRatios:double}):
\begin{equation}
  \operatorname{double~ratio}\yieldsRatioQty =
    \frac{\yieldsRatioQty_\text{\PbPb}}{\yieldsRatioQty_\text{\pp}} \,,
\end{equation}
Clearly \phi production is
enhanced roughly threefold for all 3 measured energies. This was already observed in \recite{bib:NA49phi2008}, in which a parametrization proposed in \recite{bib:E917_AuAu11_phi_2004} of the \phi production cross-section had been used as reference instead of experimental \pp data.
\par
The strangeness enhancement of \phi mesons can be compared to that of charged kaons relative to charged pions (see \figref{fig:YieldRatios:double}). It is systematically larger for \phi mesons than for kaons, being however comparable to that for positive kaons. 

\begin{figure*}[tp]
  \centering
  \stdMoveLeft
  \subfloat[]{
    \label{fig:TotalYields}
    \includegraphics[width=0.5\textwidth,page=1]{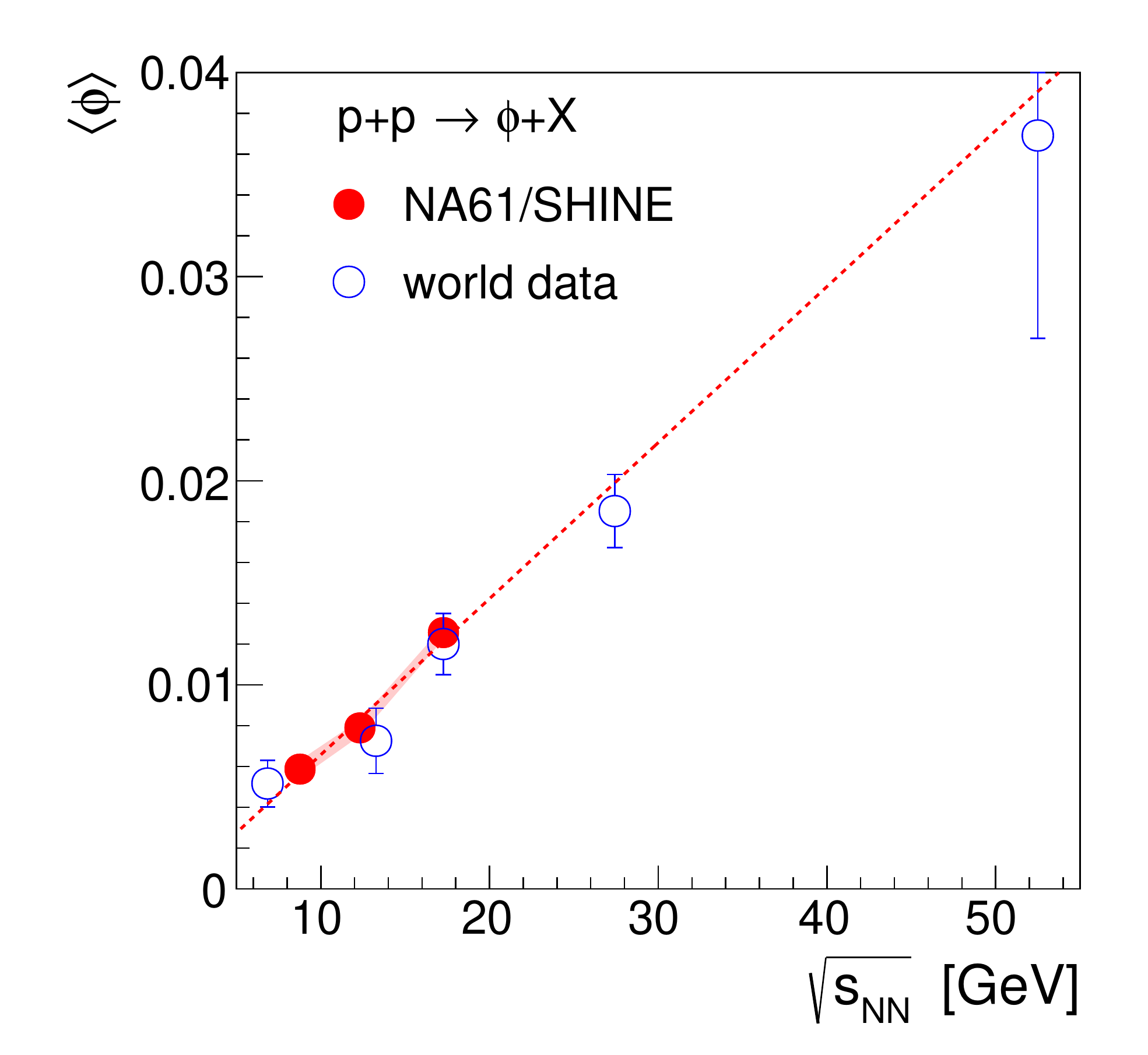}
  }
  \subfloat[]{
    \label{fig:MidrapYields}
    \includegraphics[width=0.5\textwidth,page=1]{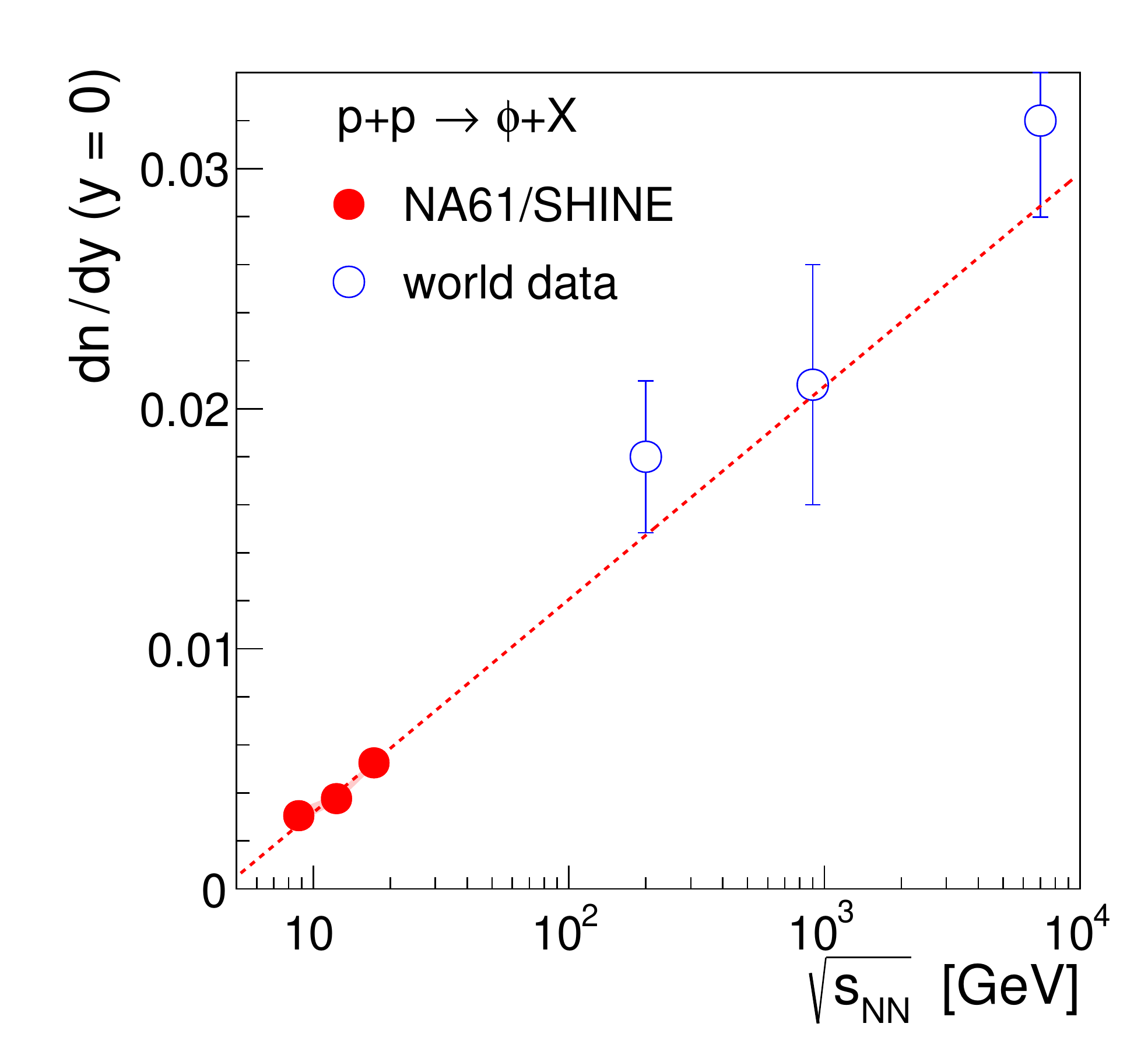}
  }
  \caption{Energy dependence of \subref{fig:TotalYields} total yields and
    \subref{fig:MidrapYields} midrapidity yields of \phi mesons in \pp
    collisions. World data on total yields come from
    \recites{bib:Blobel_pp24_phi_1975, bib:ACCMOR_hadrons63_93_phi_1981,
      bib:Drijard_ppS53_phi_1981, bib:AguilarBenitez_pp400_phi_1991,
    bib:NA49phi2000}, while on midrapidity yields come from
    \recites{bib:STAR_many_reactions_phi_2009, bib:ALICE_phi_0.9TeV,
    bib:ALICE_phi_7TeV}. Red dashed lines are fits to guide the eye (see text).
  }
  \label{fig:Yields}
\end{figure*}
\begin{figure*}[tp]
  \centering
  \stdMoveLeft
  \subfloat[]{
    \label{fig:YieldsZoom:Total}
    \includegraphics[width=0.5\textwidth,page=2]{total_yields.pdf}
  }
  \subfloat[]{
    \label{fig:YieldsZoom:Midrap}
    \includegraphics[width=0.5\textwidth,page=2]{midrap_yields.pdf}
  }
  \caption{Energy dependence of \subref{fig:YieldsZoom:Total} total yields and
    \subref{fig:YieldsZoom:Midrap} midrapidity yields of \phi mesons in \pp
    collisions at SPS energies. Also shown are the results of microscopic model calculations
    (\EPOS, \Pythia, \UrQMD) as well as the hadron resonance gas model (\HRG).
  }
  \label{fig:YieldsZoom}
\end{figure*}

Next the excitation function of \phi meson production will be discussed. 
\Figref{fig:Yields} shows the energy dependence of total and midrapidity yields of \phi mesons produced in \pp collisions. For CERN SPS and ISR energies total inclusive cross-sections are given in \recites{bib:Blobel_pp24_phi_1975,
bib:ACCMOR_hadrons63_93_phi_1981, bib:Drijard_ppS53_phi_1981,
bib:AguilarBenitez_pp400_phi_1991, bib:NA49phi2000}. They are converted into multiplicities for \Figref{fig:TotalYields} according to \equref{eq:xsectConversion}. 
At RHIC and LHC only midrapidity yields are measured~\cite{bib:STAR_many_reactions_phi_2009, bib:ALICE_phi_0.9TeV, bib:ALICE_phi_7TeV}. The corresponding excitation function is shown in \Figref{fig:MidrapYields}.
Wherever systematic uncertainties of world data are available, they are summed quadratically with statistical uncertainties for brevity of presentation.
\par
Straight lines are fitted to the data points in
\figref{fig:TotalYields} assuming proportionality between the total energy available for production and the number of produced \phi mesons. All measurements, i.e.\ world data and those from this analysis, are used in the fit. 
The resulting straight (red) line describes well the data in the considered energy range. 
\par
For midrapidity yields no well-motivated parameterisation of \sNN dependence exists. For simplicity the red dashed line in \figref{fig:MidrapYields} corresponds to the function
\begin{equation}
  f\pqty{\sNN} = a \log_{10}\pqty{\sNN / b} \,,
\end{equation}
which is a guess and happens to describe well the data points. 
\par
While \Figref{fig:Yields} covers the energy range up to LHC energies \Figref{fig:YieldsZoom} zooms in on the SPS energy range for the comparison to model calculations. One observes that the slope of the excitation function is reproduced by \Pythia, \UrQMD, and \EPOS within statistical and systematic uncertainties for both the total and mid-rapidity yields. The yields, however are off by factors of 0.25 and 0.7 for \Pythia/\Epos and \UrQMD, respectively. The hadron gas model (\HRG~\cite{bib:Vovchenko2016}) overpredicts the total yields by roughly a factor of two.

\begin{figure*}[!t]
  \centering
  \stdMoveLeft
  \subfloat[]{
    \label{fig:SigmaYvsYbeam}
    \includegraphics[width=0.5\textwidth,page=3]{sigma_y.pdf}
  }
  \subfloat[]{
    \label{fig:Coalescence}
    \includegraphics[width=0.5\textwidth,page=4]{sigma_y.pdf}
  }
  \caption{\subref{fig:SigmaYvsYbeam} Widths of rapidity distributions of
    various particles in \pp (full symbols) and central \PbPb collisions (open
    symbols) as a function of beam rapidity.
    Full red circles are results of this analysis, the star is the \pp NA49 measurement~\cite{bib:NA49phi2000}, other \pp points come from
    \NASixtyOne~\cite{bib:NA61_pp_pion_2014, bib:PulawskiPhD}, while the \PbPb points
    from NA49~\cite{bib:NA49phi2008, bib:NA49Onset2002, bib:NA49Onset2008,
    bib:NA49Lambdas2004}. Lines are fitted to points to guide the eye.
    \subref{fig:Coalescence} Comparison of widths for \phi mesons with
    expectations from kaon coalescence (see text) and models.
  }
\end{figure*}

The last paragraph of this section addresses the so far not explained exceptional role which the \phi meson plays when considering the widths of the rapidity distributions (\sigmaY) of produced particles as function of energy put forward by the NA49 collaboration~\cite{bib:NA49phi2008}. Except for the \phi these widths increase approximately linearly with beam rapidity, at the same rate and irrespective of the colliding system.
\Figref{fig:SigmaYvsYbeam} shows the widths of the rapidity distributions of \phi mesons and various other produced particles in \pp and central \PbPb collisions as a function of beam rapidity in the centre-of-mass frame. The corresponding figure in \recite{bib:NA49phi2008} has been  complemented by the \NASixtyOne results on \pim~\cite{bib:NA61_pp_pion_2014} and \Kp,\Km in \pp collisions. The \sigmaY of \Kp and \Km were calculated from the distributions given in~\recite{bib:PulawskiPhD}. 
The excitation function of \sigmaY for the \phi meson in \PbPb collisions is significantly steeper than the one of the other particles. The peculiarity of this result is emphasized by the new \NASixtyOne \pp data: the \phi data points suggest that it is not the \phi meson which is peculiar in itself, it rather is something specific to \phi meson production in the heavy \PbPb system.
Kaon coalescence is a possible source of \phi mesons in the final state. It correlates \sigmaY of \phi with \sigmaY of kaons. Thus one can calculate \sigmaY of \phi from \sigmaY of kaons in \pp and \PbPb using the method described in \recite{bib:NA49phi2008}. The result is shown as thick
black lines for \pp (solid) and \PbPb (dotted) in \figref{fig:Coalescence} 
together with the experimental data. The difference between the coalescence expectations and the actual measurements is much smaller for \pp than for \PbPb data points.

\section{Summary and conclusions}\label{s:summary}
Spectra and multiplicities of \phi mesons produced in inelastic p+p interactions were measured with the NA61/SHINE spectrometer at beam momenta \SI{40}, \SI{80}, \SI{158}{\GeVc} at the CERN SPS. The tag-and-probe method, adapted from LHC analyses, was used to analyze the \Kp - \Km invariant mass distributions. For the \SI{158}{\GeVc} and \SI{80}{\GeVc} data sets the analysis was done double differentially yielding spectra of rapidity and transverse momentum. 
The limited number of \phi candidates for \SI{40}{\GeVc} allowed only for a single differential analysis resulting in transverse momentum and rapidity spectra integrated over rapidity and transverse momentum, respectively. The statistical errors are larger than the systematic uncertainties for all energies. While each of the considered microscopic models reproduces the shape of either the transverse momentum or the rapidity spectra, none describes both consistently. 
\par
\NASixtyOne results on \phi production in \pp collisions are the elementary reference for the study of collective effects in \PbPb data~\cite{bib:NA49phi2008}. They emphasize the intriguing energy dependence of \sigmaY of the \phi meson in central \PbPb collisons. The widths of rapidity spectra in \pp and \PbPb collisions are systematically larger than expected from the hypothesis that \phi mesons are predominantly produced through kaon coalescence. Kaon coalescence can still be the most important mechanism in \pp interactions, however in \PbPb collisions a new production process for \phi mesons seems to become important at higher energies, which is not present in pion, kaon, and anti-Lambda production.
Our findings at \SI{158}{\GeVc} agree with previously published results from the NA49 collaboration~\cite{bib:NA49phi2000} within quoted uncertainties. The latter are almost 3 times smaller in the \NASixtyOne than in the NA49 data. Neither total yields nor spectra on \phi production in \pp interactions have previously been published at beam energies of \SI{40}{\GeVc} and \SI{80}{\GeVc}.
\par
Our results confirm that the excitation function of \phi multiplicity is almost perfectly linear in \pp interactions. In the low energy regime neither the three microscopic models, \recite{bib:EPOS2006, bib:EPOS2009, bib:PythiaManual, bib:UrQMD1998, bib:UrQMD1999} nor the statistical hadron gas model~\cite{bib:Vovchenko2016} can reproduce the experimmental excitation function quantitatively, the precision of which was increased  significantly by the \NASixtyOne results.


\section*{Acknowledgments}
We would like to thank the CERN EP, BE and EN Departments for the
strong support of NA61/SHINE.
\IfEPJC{}{\par}
This work was supported by
the Hungarian Scientific Research Fund (grant NKFIH 123842\slash123959),
the Polish Ministry of Science
and Higher Education (grants 667\slash N-CERN\slash2010\slash0,
NN 202 48 4339 and NN 202 23 1837), the National Science Centre Poland (grants 2011\slash03\slash N\slash ST2\slash03691,
2013\slash11\slash N\slash ST2\slash03879, 
2014\slash13\slash N\slash ST2\slash02565, 
2014\slash14\slash E\slash ST2\slash00018,
2014\slash15\slash B\slash ST2\slash02537,
2015\slash18\slash M\slash ST2\slash00125, 
2015\slash 19\slash N\slash ST2\slash01689, 
2016\slash23\slash B\slash ST2\slash00692, 
2017\slash 25\slash N\slash ST2\slash 02575,
2018\slash 30\slash A\slash ST2\slash 00226),
the Russian Science Foundation, grant 16-12-10176, 
the Russian Academy of Science and the
Russian Foundation for Basic Research (grants 08-02-00018, 09-02-00664
and 12-02-91503-CERN), the Ministry of Science and
Education of the Russian Federation, grant No.\ 3.3380.2017\slash4.6,
 the National Research Nuclear
University MEPhI in the framework of the Russian Academic Excellence
Project (contract No.\ 02.a03.21.0005, 27.08.2013),
the Ministry of Education, Culture, Sports,
Science and Tech\-no\-lo\-gy, Japan, Grant-in-Aid for Sci\-en\-ti\-fic
Research (grants 18071005, 19034011, 19740162, 20740160 and 20039012),
the German Research Foundation (grant GA\,1480/8-1), the
Bulgarian Nuclear Regulatory Agency and the Joint Institute for
Nuclear Research, Dubna (bilateral contract No. 4799-1-18\slash 20),
Bulgarian National Science Fund (grant DN08/11), Ministry of Education
and Science of the Republic of Serbia (grant OI171002), Swiss
Nationalfonds Foundation (grant 200020\-117913/1), ETH Research Grant
TH-01\,07-3 and the U.S.\ Department of Energy.

\bibliographystyle{na61Utphys}
\bibliography{includes/bibliography}

\appendix

\newpage
{\Large The \NASixtyOne Collaboration}
\bigskip
\begin{sloppypar}

\noindent
A.~Aduszkiewicz$^{\,15}$,
E.V.~Andronov$^{\,21}$,
T.~Anti\'ci\'c$^{\,3}$,
V.~Babkin$^{\,19}$,
M.~Baszczyk$^{\,13}$,
S.~Bhosale$^{\,10}$,
A.~Blondel$^{\,23}$,
M.~Bogomilov$^{\,2}$,
A.~Brandin$^{\,20}$,
A.~Bravar$^{\,23}$
% ...

\end{sloppypar}

\noindent
% ...
$^{2}$~Faculty of Physics, University of Sofia, Sofia, Bulgaria\\
$^{3}$~Ru{\dj}er Bo\v{s}kovi\'c Institute, Zagreb, Croatia\\
% ...
$^{10}$~Institute of Nuclear Physics, Polish Academy of Sciences, Cracow, Poland\\
% ...
$^{13}$~AGH - University of Science and Technology, Cracow, Poland\\
% ...
$^{15}$~University of Warsaw, Warsaw, Poland\\
% ...
$^{19}$~Joint Institute for Nuclear Research, Dubna, Russia\\
$^{20}$~National Research Nuclear University (Moscow Engineering Physics Institute), Moscow, Russia\\
$^{21}$~St. Petersburg State University, St. Petersburg, Russia\\
% ...
$^{23}$~University of Geneva, Geneva, Switzerland\\
% ...

\end{document}